%% file: ANA-EXOT-2019-20-PAPER.tex
\pdfoutput=1
\documentclass[cernpreprint, atlasdraft=false, UKenglish, texmf, orcidlogo]{atlasdoc}
 
\usepackage{atlaspackage}
\usepackage{atlasbiblatex}
 
\usepackage[BSM=true]{atlasphysics}

\RequirePackage{scrlfile}
\ReplacePackage{scrpage2}{scrlayer-scrpage}
\usepackage{multirow}
\usepackage{longtable}
\usepackage{siunitx}
\usepackage{comment}
\usepackage{caption}
\DeclareCaptionStyle{capstyle}[justification=centering]{format=hang, labelfont={rm, bf}, textfont={rm} }
\graphicspath{{logos/}{figs/}}
 
\usepackage{ANA-EXOT-2019-20-PAPER-defs}
 
\mathchardef\mhyphen="2D

\AtlasTitle{Search for lepton-flavour violation in high-mass dilepton
final states using 139 $\mathrm{fb}^{-1}$ of $pp$ collisions at $\sqrt{s}=13$ TeV
with the ATLAS detector}
 
\usepackage{authblk}

\AtlasAbstract{
A search is performed for a heavy particle decaying into different-flavour,
dilepton final states, using 139 $\mathrm{fb}^{-1}$
of proton--proton collision data at $\sqrt{s}=13$ TeV collected in
2015--2018 by the ATLAS detector at the Large Hadron Collider. Final states with electrons, muons and
hadronically decaying tau leptons are considered ($e\mu$, $e\tau$ or $\mu\tau$).
No significant excess over the Standard Model predictions is observed.
Upper limits on the production cross-section are set as a function of the mass of a Z' boson, a supersymmetric $\tau$-sneutrino, and a quantum black-hole. The observed 95\% CL lower mass limits obtained on a typical benchmark model Z' boson are 5.0 TeV (e$\mu$), 4.0 TeV (e$\tau$), and 3.9 TeV ($\mu\tau$), respectively.} 
 
\author{The ATLAS Collaboration}
 
\AtlasRefCode{EXOT-2019-20}
 
\PreprintIdNumber{CERN-EP-2023-089}
 
\AtlasDate{\today}

\AtlasJournal{JHEP}
\AtlasJournalRef{J. High Energ. Phys. 2023, 82 (2023).}
\AtlasDOI{10.1007/JHEP10(2023)082}

\hypersetup{pdftitle={ATLAS draft},pdfauthor={The ATLAS Collaboration}}
 
\begin{document}
 
\maketitle

\clearpage
 
\section{Introduction}

Lepton-flavour-violation (LFV) is forbidden in the Standard Model (SM) theory of particle physics. However,
the observation of flavour oscillations among neutrinos has shown that lepton flavour is not a conserved
symmetry of nature~\cite{PhysRevLett.81.1562}. The search for charged LFV (CLFV)
remains an area of active interest, motivating many experimental searches for the CLFV decays, such as the MEG experiment for $\mu\to e\gamma$~\cite{themegcollaboration2016search}, $\mu\to 3e$ with the SINDRUM spectrometer~\cite{BELLGARDT19881} and also many searches of LFV $\tau$-lepton decay~\cite{Amhis_2021}. So far, there is no experimental evidence that LFV also occurs in interactions
between charged leptons, and any observation of LFV would be a clear signature of new physics.
 
Many extensions of the SM predict LFV couplings, such as models with \Zprime bosons \cite{bib:Langacker2008}, scalar neutrinos in
$R$-parity-violating (RPV) \cite{Farrar:1978xj,bib:RParitySUSY} supersymmetry (SUSY) and
quantum black holes (QBH) in low-scale gravity \cite{bib:Gingrich2009}.
These dilepton LFV processes are usually produced at TeV energy scale, with a clear detector signature of a
prompt, different-flavour lepton pair, which decreases the contribution from SM background processes.
 
A common extension of the SM is the addition of an extra $U(1)$ gauge symmetry
resulting in a massive neutral vector boson known as a \Zprime boson.
The Sequential Standard Model (SSM)~\cite{bib:Langacker2008} is used as a benchmark in this paper,
where the \Zprime boson is assummed to have the same quark couplings and chiral structure as the SM $Z$ boson, but allowing for lepton flavour violating couplings.
Additonal LFV coupling parameters $Q_{ij}$ are
assigned for $\Zprime\to{\ell_i}{\ell_j}$ processes, where $i,j = 1 \ldots 3$ represent the
three lepton generations. It is assumed that the $Q_{ij}$ parameters equal to the SM $Z$ boson coupling for $i=j$.
The latest ATLAS Collaboration study placed lower limits of 4.5,
3.7, and \SI{3.5}{\TeV} on the mass of a \Zprime boson decaying into $e\mu$,
$e\tau$, and $\mu\tau$ pairs, respectively,
using \SI{36.1}{\per\femto\barn} of the \SI{13}{\TeV} data sample
\cite{EXOT-2016-09}. The CMS Collaboration has placed limits up to
\SI{5.0}{\TeV} on a \Zprime boson decaying into an $e\mu$ final
state using \SI{138}{\per\femto\barn} \cite{CMS-EXO-19-014}. Following the
same methodology as in Ref.~\cite{EXOT-2015-20},
the polarisation of $\tau$-leptons is not included in the model, but its
impact on the sensitivity to a possible signature is found to be negligible.

In the RPV SUSY model, the Lagrangian terms allowing LFV can be expressed as
$$\frac{1}{2}\lambda_{ijk}L_{i}L_{j}\bar{e}_k + \lambda^{\prime}_{ijk}L_{i}Q_{j}\bar{d}_k,$$
where $L$ and $Q$ are the $SU(2)$ doublet superfields of leptons and quarks,
$e$ and $d$ are the $SU(2)$ singlet superfields of charged leptons and
down-type quarks, $\lambda$ and $\lambda^\prime$ are Yukawa couplings, and the
indices $i$, $j$, and $k$ denote generations. A $\tau$-sneutrino
($\tilde{\nu}_{\tau}$) may be produced in proton--proton ($pp$) collisions by $d\bar{d}$
annihilation and later decay into $e\mu$, $e\tau$, or $\mu\tau$. Although
only $\tilde{\nu}_{\tau}$ is considered in this paper, results apply to any
sneutrino flavour. For the theoretical prediction of the cross-section times
branching ratio, the $\tilde{\nu}_{\tau}$ coupling to first-generation quarks
($\lambda^{\prime}_{311}$) is assumed to be 0.11 for all channels. As in the
\Zprime model, each lepton-flavour-violating final state is considered separately.
It is assumed that $\lambda_{312} = \lambda_{321} = 0.07$ for the $e\mu$ final state,
$\lambda_{313} = 0.07$ for the $e\tau$ final state, and $\lambda_{323} = 0.07$ for the $\mu\tau$ final state, while $\lambda_{331}$ and $\lambda_{332}$
are set to be zero, due to the gauge invariance for these channels, resulting in the $\tilde{\nu}_{\tau}$ cross-section
times branching ratio in the $e\mu$ channel being up to approximately twice as large as in the $e\tau$ or $\mu\tau$ channel. These values are chosen for easy comparisons with previous ATLAS and CMS searches
\cite{EXOT-2015-20,SUSY-2013-06,CMS-EXO-13-002}.
However in the previous ATLAS search~\cite{EXOT-2016-09} and the latest CMS publication~\cite{CMS-EXO-19-014}, $\lambda_{331}$ and $\lambda_{332}$ were assumed to be the same as $\lambda_{313}$ and $\lambda_{323}$ respectively, so the results of RPV model in this paper are not directly comparable to those in the previous publications for $e\tau$ and $\mu\tau$ channels.
The latest ATLAS results with \SI{36.1}{\per\femto\barn} using the \SI{13}{\TeV} data~\cite{EXOT-2016-09}
have excluded RPV SUSY models  below the $\tilde{\nu}_{\tau}$ masses
\SI{3.4}{\TeV}, \SI{2.6}{\TeV} and below  \SI{2.3}{\TeV} for $e\mu$, $e\tau$ and $\mu\tau$
channels respectively with exactly the same parameters mentioned above. The CMS
Collaboration has recently excluded 
RPV SUSY models below
\SI{2.2}{\TeV} for $\lambda_{312} = \lambda_{321} = \lambda^{\prime}_{311} = 0.01$~\cite{CMS-EXO-19-014}.
 
Various models introduce extra spatial dimensions to reduce the value of the
Planck mass and resolve the hierarchy problem. The search described in this
paper presents interpretations based on two models: the Arkani-Hamed--Dimopoulos--Dvali
(ADD) model~\cite{bib:Hamed1998}, assuming $n=6$, where $n$ is the number of
extra dimensions, and on the Randall--Sundrum (RS) model \cite{bib:Randall1999}
with one extra dimension. Due to the increased strength of gravity at short
distances, in these models $pp$ collisions at the LHC could produce states
exceeding the threshold mass ($m_{\mathrm{th}}$) to form black holes.
In this paper, $m_{\mathrm{th}}$ is assumed to be equivalent to the
extra-dimensional Planck scale. The quantum gravity regime \cite{bib:MeadeRandall2007,
bib:Gingrich2008,bib:Calmet2008} is applied only when considering the mass region below
3--5 $m_{\mathrm{th}}$, since for masses beyond this region it is expected that
thermal black holes would be produced. The non-thermal (or quantum) black holes
could decay into two-particle final states, producing the topology investigated in this paper.
QBHs would have a continuum mass distribution from
$m_{\mathrm{th}}$ up to the beginning of the thermal regime which for
the models considered in this paper is assumed to start at
$3 m_{\mathrm{th}}$. This approach is consistent with the previous ATLAS analyses, such as Ref.\cite{EXOT-2016-09}.
The decay of the QBH would be governed by a yet
unknown theory of quantum gravity. The two main assumptions of the
extra-dimension models considered in this paper \cite{bib:Gingrich2009} are  that (a) gravity couples with equal strength to all SM particle degrees of freedom
and (b) gravity conserves local symmetries (colour and electric charge), but can
violate global symmetries such as lepton-flavour and baryon-number conservation.
Following these assumptions, the branching ratio to each final state
can be calculated. QBHs decaying into different-flavour, opposite-charge lepton pairs
are created via $q\bar{q}$ ($gg$) with a branching ratio to $\ell\ell^\prime$ of 0.87\% (0.34\%) \cite{bib:Gingrich2009}. As in the
\Zprime model, each lepton-flavour-violating final state is considered separately. These models were used in
previous ATLAS and CMS searches for QBH in dijet
\cite{EXOT-2013-11,EXOT-2015-02,CMS-EXO-12-059}, lepton+jet
\cite{EXOT-2013-03}, photon+jet \cite{EXOT-2012-22}, $e\mu$
\cite{CMS-EXO-19-014}, and same-flavour dilepton \cite{EXOT-2012-23} final
states. The latest ATLAS Collaboration study placed lower limits of 5.5 (3.4),
4.9 (2.9), and 4.5 (2.6) \SI{}{\TeV} on $m_{\mathrm{th}}$ of QBH with ADD (RS) model decaying into $e\mu$,
$e\tau$, and $\mu\tau$ pairs, respectively, using \SI{36.1}{\per\femto\barn} of the \SI{13}{\TeV} data sample
\cite{EXOT-2016-09}.

This paper describes a search for new phenomena in final states with two leptons of
different flavour using \SI{139}{\per\femto\barn} of data from $pp$ collisions
at $\sqrt{s}=13$~\si{\TeV} at the Large Hadron Collider (LHC). The dilepton signal final states
consisting of $e\mu$, $e\tau$, or $\mu\tau$ pairs are considered, where the $\tau$-lepton decays hadronically.
This analysis is looking for a localised excess in the distribution of dilepton invariant mass in TeV range.
There are three signal regions defined, one for each decay mode. Corresponding control regions are also designed to extract a normalisation of the most prominent SM backgrounds: production of top quarks and dibosons. 
The contributions from fake leptons are calculated from the data. The final simultaneous fit, with the signal and control regions, is performed separately in each decay mode. Four benchmark models (\Zprime, RPV SUSY $\tilde{\nu}_{\tau}$ and QBH: ADD n=6; RS n=1) are used to interpret the results.
Compared with the previous ATLAS search with \SI{36}{\per\femto\barn} of $pp$
collision data at $\sqrt{s}=13$~\si{\TeV}~\cite{EXOT-2016-09}, this analysis benefits from a factor of four increase
in integrated luminosity, improvements in object reconstruction (such as electron and tau lepton identification) the
use of a more robust background estimation method (e.g. the 4 $\times$ 4 matrix method in $e\mu$ channel), and the application of
a simultaneous fit with both signal region and control regions to constrain systematic uncertainties. In this analysis, the b-jet veto is also included as a part of the baseline selection in every channel, which highly suppresses the top quark related backgrounds.


\section{The ATLAS detector}

The ATLAS detector \cite{PERF-2007-01} is a general-purpose particle
detector with approximately forward--backward symmetric cylindrical
geometry. It covers nearly the entire solid angle around the collision point.\footnote{
ATLAS uses a right-handed coordinate system with its origin at the nominal interaction point (IP)
in the centre of the detector and the \(z\)-axis along the beam pipe.
The \(x\)-axis points from the IP to the centre of the LHC ring,
and the \(y\)-axis points upwards.
Cylindrical coordinates \((r,\phi)\) are used in the transverse plane,
\(\phi\) being the azimuthal angle around the \(z\)-axis.
The pseudorapidity is defined in terms of the polar angle \(\theta\) as \(\eta = -\ln \tan(\theta/2)\).
Angular distance is measured in units of \(\Delta R \equiv \sqrt{(\Delta\eta)^{2} + (\Delta\phi)^{2}}\).
}
It consists of an inner tracking detector surrounded by a thin superconducting solenoid, electromagnetic and hadron calorimeters,
and a muon spectrometer incorporating three large superconducting air-core toroidal magnets.
 
The inner-detector system (ID) is immersed in a \qty{2}{\tesla} axial magnetic field
and provides charged-particle tracking in the range of \(|\eta| < 2.5\).
The high-granularity silicon pixel detector covers the vertex region and typically provides four measurements per track,
the first hit normally being in the insertable B-layer (IBL) installed before Run~2~\cite{ATLAS-TDR-19,PIX-2018-001}.
It is followed by the silicon microstrip tracker (SCT), which usually provides eight measurements per track.
These silicon detectors are complemented by the transition radiation tracker (TRT),
which enables radially extended track reconstruction up to \(|\eta| = 2.0\).
The TRT also provides electron identification information
based on the fraction of hits (typically 30 in total) above a higher energy-deposit threshold corresponding to transition radiation.
 
The calorimeter system covers the pseudorapidity range \(|\eta| < 4.9\).
Within the region of \(|\eta|< 3.2\), electromagnetic calorimetry is provided by barrel and
endcap high-granularity lead/liquid-argon (LAr) calorimeters,
with an additional thin LAr presampler covering \(|\eta| < 1.8\)
to correct for energy loss in material upstream of the calorimeters.
Hadron calorimetry is provided by the steel/scintillator-tile calorimeter,
segmented into three barrel structures within \(|\eta| < 1.7\), and two copper/LAr hadron endcap calorimeters.
The solid angle coverage is completed with forward copper/LAr and tungsten/LAr calorimeter modules
optimised for electromagnetic and hadronic energy measurements respectively.
 
The muon spectrometer (MS) comprises separate trigger and
high-precision tracking chambers measuring the deflection of muons in a magnetic field generated by the superconducting air-core toroidal magnets.
The field integral of the toroids ranges between \num{2.0} and \qty{6.0}{\tesla\metre}
across most of the detector.
Three layers of precision chambers, each consisting of layers of monitored drift tubes, cover the region \(|\eta| < 2.7\),
complemented by cathode-strip chambers in the forward region, where the background is highest.
The muon trigger system covers the range \(|\eta| < 2.4\) with resistive-plate chambers in the barrel, and thin-gap chambers in the endcap regions.
 
Interesting events are selected by the first-level trigger system implemented in custom hardware,
followed by selections made by algorithms implemented in software in the high-level trigger~\cite{TRIG-2016-01}.
The first-level trigger accepts events from the \qty{40}{\MHz} bunch crossings at a rate below \qty{100}{\kHz},
which the high-level trigger further reduces in order to record events to disk at about \qty{1}{\kHz}.
 
An extensive software suite~\cite{ATL-SOFT-PUB-2021-001} is used in data simulation, in the reconstruction
and analysis of real and simulated data, in detector operations and in the trigger and data acquisition
systems of the experiment.


\section{Data and simulated event samples}

The data sample used for this analysis was collected during 2015 to
2018 from $pp$ collisions at a centre-of-mass energy of \SI{13}{\TeV}. After
selecting periods with stable beams and applying data-quality requirements, the
total integrated luminosity is \SI{139}{\per\femto\barn} with an uncertainty of 1.7\%~\cite{DAPR-2021-01,LUCID2}.
 
The $pp\to\Zprime\to\ell\ell^{\prime}$ Monte Carlo (MC) simulated samples were generated at leading
order (LO) using the generator \Pythia~8.186~\cite{bib:Pythia8}
with the NNPDF2.3LO \cite{bib:Ball2012} parton distribution function (PDF) set
and the A14~\cite{ATL-PHYS-PUB-2014-021} set of tuned parameters. The signal samples were generated for the
SSM, including
fifteen mass points ranging from \SI{1}{\TeV} to \SI{8}{\TeV} in
steps of \SI{500}{\GeV}. The cross-section of the \Zprime signal was corrected from LO to next-to-next-to-leading order (NNLO) in the strong coupling constant with a rescaling, which was computed
with VRAP 0.9 \cite{Anastasiou:2003ds} and the CT14NNLO PDF \cite{bib:Dulat2015} set. The NNLO QCD correction was applied as a multiplicative factor from 1.44 to 0.29 for different \Zprime masses on total cross-section. No mixing of the $Z^\prime$ boson with the $Z$ and $\gamma^*$ bosons is included.
 
The $d\bar{d}\to\tilde{\nu}_{\tau}\to\ell\ell^{\prime}$ samples were
generated at LO with \Madgraph~v2.3.3~\cite{bib:MadGraph5}
interfaced to the \Pythia~8.186 parton shower model with the NNPDF2.3LO PDF set
and the A14 tune. The signal samples were generated with the
same mass values as for the \Zprime model described above. The cross-section was
calculated at LO with the same generator used for simulation and corrected
to next-to-leading-order (NLO) using \LoopTools~v2.2~\cite{bib:LoopTools}.
The NLO correction factor ranges from 1.1 to 1.4, depending on different
$\tilde{\nu}_{\tau}$ masses.
 
The $pp\to$ QBH  $\to\ell\ell^{\prime}$ samples were generated with the program \QBH~3.00
\cite{bib:QBHManual}  using the CTEQ6L1 \cite{bib:CTEQ6L1} PDF set and the A14
tune, for which \Pythia~8.205~\cite{Sjostrand:2014zea} provides showering and
hadronisation. For each extra-dimensional model, eleven $m_{\mathrm{th}}$ points
in \SI{500}{\GeV} steps were produced: from \SI{4}{\TeV} to \SI{9}{\TeV} for the ADD $n=6$
model and from \SI{2}{\TeV} to \SI{7}{\TeV} for the RS $n=1$ model. 
These two models differ in the number and nature of the additional extra
dimensions (large extra dimensions for ADD and one highly warped extra
dimension for RS). In particular, the ADD model predicts black
holes with a larger gravitational radius and hence the parton--parton
cross-section for this model is larger than for the RS model. Therefore, the
$m_{\mathrm{th}}$ range of the generated samples differs for the two models.
 
The SM background in the LFV dilepton search is due to several processes
which produce a final state with two different-flavour leptons. For the $e\mu$
mode, the dominant background contributions originate from \ttbar and single-top $t\W$ production
with the subsequent decay of both of the \W~bosons (including those from top quark decays) in the event into leptons, and diboson
(\WW, \WZ, and \ZZ) production. Other backgrounds originate from $\tau$-lepton pair production
($q\bar{q} \to\Z/\gamma^{*}\to\tau\tau$). Both diboson and $\tau$-lepton pair production produce
different-flavour
final states, through the leptonic decays of the \W~and \Z~bosons or the $\tau$-leptons.
For the $e\tau$ and $\mu\tau$ modes, multijet and \W+jets processes are the dominant backgrounds. They
contribute due to the misidentification of jets as leptons.

Backgrounds from top quark production include \ttbar and \(tW\) production. The production of \ttbar events was modelled using the
\powhegbox~v2~\cite{Frixione:2007nw,Nason:2004rx,Frixione:2007vw,Alioli:2010xd}
generator at NLO with the NNPDF3.0NLO PDF set~\cite{Ball:2014uwa}
and the \hdamp parameter\footnote{The
\hdamp parameter is a resummation damping factor and one of the
parameters that controls the matching of \POWHEG matrix elements to
the parton shower and thus effectively regulates the
high-\pT radiation against which the \ttbar system recoils.} set
to 1.5\,\ensuremath{m_{\text{top}}}\xspace~\cite{ATL-PHYS-PUB-2016-020}.  The events were interfaced
to \Pythia~8.230~\cite{Sjostrand:2014zea} to model the parton shower,
hadronisation, and underlying event, with parameters set according
to the A14 tune and using the NNPDF2.3LO
set of PDFs. The decays of bottom and charm hadrons
were performed by \EVTGEN[1.6.0]~\cite{Lange:2001uf}.
The \(tW\) backgrounds were
modelled by the
\powhegbox~v2 generator at NLO in QCD using the five-flavour scheme and the
NNPDF3.0NLO set of PDFs.
The diagram removal scheme~\cite{Frixione:2008yi} was used to
remove interference and overlap with \ttbar production. The events were interfaced to \Pythia~8.230 using the A14
tune and the NNPDF2.3LO set of
PDFs.

For diboson samples,
final states were simulated with \textsc{Sherpa}\,2.2.12~\cite{Gleisberg:2008ta,Hoeche:2009rj},
including off-shell effects and Higgs boson contributions where appropriate.
Fully leptonic final states and semileptonic final states, where one boson decays leptonically and the other hadronically,
were simulated using matrix elements (ME) at NLO accuracy in QCD for up to one additional
parton and at LO accuracy for up to three additional parton emissions.

The SM Drell--Yan process was generated at NLO using \powhegbox~v1 MC generator which is
used for the simulation at NLO accuracy of the hard-scattering processes of
$Z$-boson production and decay in the $e$, $\mu$, and $\tau$-lepton final states. It was interfaced to \Pythia~8.186
with parameters set according to the \AZNLO~\cite{STDM-2012-23}
tune. The CT10 PDF set~\cite{bib:Lai2010} was used for the hard-scattering processes,
whereas the CTEQ6L1 PDF set was used for the parton shower.

Processes such as \W+jets and multijet production with jets that are
misidentified as leptons were estimated through a combination of data-driven
methods and simulation, detailed in Section~\ref{sec:background}. The \W+jets
contribution was estimated with the aid of the \Sherpa~2.2.2 simulated samples using NLO MEs for up to two partons,
and LO matrix elements for up to four partons calculated with the Comix~\cite{Gleisberg:2008fv} and \openloops~libraries. They
are matched with the \Sherpa~parton shower using the MEPS@NLO~\cite{Hoeche:2012yf} prescription using the set of
tuned parameters developed by the \Sherpa~authors~\cite{Gleisberg:2008ta}.
 
For all samples used in this analysis, the effects of multiple proton--proton
interactions per bunch crossing (pileup) were included by overlaying
minimum-bias events simulated with \Pythia~8.186 and reweighting the simulated
events to reproduce the distribution of the number of interactions per bunch
crossing observed in the data. The generated events were processed with the
ATLAS simulation infrastructure~\cite{SOFT-2010-01}, based on
\geant4~\cite{bib:Geant4}, and passed through the trigger simulation and the
same reconstruction software used for the data.


\section{Event reconstruction and selection}

The search for new phenomena presented here is aimed at the high mass range, thus
event selections are optimised accordingly.
Events are kept only if they can satisfy a single-muon or single-electron trigger.
The trigger $p_{\mathrm{T}}$ threshold is different for different data-taking periods.
For 2015 data runs, a $p_{\mathrm{T}}$ threshold of 20 or \SI{50}{\GeV} for
muons is applied, and the value for electrons is 24, 60 or \SI{120}{\GeV}.
For 2016 - 2018 data runs, the $p_{\mathrm{T}}$ threshold is 26 or
\SI{50}{\GeV} for muons, and 26, 60 or \SI{140}{\GeV} for electrons. This analysis relies on triggers with $p_{\mathrm{T}}$ threshold above 50 (60) \GeV for muons (electrons).
The single-electron trigger with higher $p_{\mathrm{T}}$ threshold has a looser
identification requirement, resulting in an increased trigger efficiency in the high $p_{\mathrm{T}}$ region which is especially important for a high mass resonance
search. In addition to the trigger selection, events are required to have at least one offline
reconstructed signal lepton matched to the object that fired the trigger.

Electron candidates are formed by associating the energy in clusters of cells
in the electromagnetic calorimeter with a track in the ID \cite{ATLAS-CONF-2014-032}.
Candidate electrons are identified using a likelihood-based method. The likelihood
discriminant utilises lateral and longitudinal calorimeter shower shapes together with tracking
and cluster--track matching quantities.
The discriminant criterion is a function of the \pT\ and $|\eta|$ of
the electron candidate. Two operating points are used in this analysis, as
defined in Ref.~\cite{ATL-PHYS-PUB-2015-041,PERF-2017-01}: \emph{Medium} and \emph{Tight},
which correspond to an efficiency of 88\% and 80\% for a prompt electron with $E_{\mathrm{T}}=40$ \GeV respectively.
The \emph{Tight} working point is required for electrons in signal events while
the \emph{Medium} working point is required in order to define an 
electron
used for the reducible background estimates.
Electron candidates must have
$\pT > 65$~\si{\GeV} and $|\eta|<2.47$, excluding the region
$1.37<|\eta|<1.52$, where the electron performance is degraded due
to the presence of extra inactive material. Further requirements are made on their tracks:
the transverse and longitudinal impact parameters relative to the
primary vertex of the event ($d_0$ and $\Delta z_0$) must satisfy
$|d_{\mathrm{0}}/\sigma_{d_{\mathrm{0}}}|<5$ and
$|\Delta z_{\mathrm{0}}\sin\theta|<0.5$~\si{\milli\meter}.
Candidates are required to satisfy relative track-based
and calorimeter-based isolation requirements with an efficiency
of 99\% to suppress background from non-prompt electrons~\cite{Hyneman:2672803}.

Candidate muon tracks are reconstructed independently in the ID and
the MS which are then used in a combined fit.
To ensure optimal muon momentum resolution  to very high $p_{\mathrm{T}}$
region ($\sim$10\% at 1~\TeV), the \emph{High}-$p_{\mathrm{T}}$ operating
point~\cite{PERF-2015-10} is used in this analysis. Only tracks with
hits in each of the three stations of the muon spectrometer are considered.
Moreover, muon candidates are required to be within a  pseudorapidity of  $|\eta| <$ 2.5,
satisfy  $|d_{\mathrm{0}}/\sigma_{d_{\mathrm{0}}}|<3$ and
$|\Delta z_{\mathrm{0}}\sin\theta|<0.5$~\si{\milli\meter},
$\pT > 65$~\si{\GeV}, and satisfy a
track-based isolation criterion with an efficiency of 99\% to further
reduce contamination from non-prompt muons. The track isolation is
similar to the one defined for electrons.
Muon candidates fulfilling all selection except the isolation criterion are
called ``Loose muons'' which are used in the reducible background estimates. An additional upper cut on $\mu$ \pT at 6 \TeV is applied to remove poorly calibrated muons. This only affects the very high-mass signals, and no events fail this requirement in either the data or the background MC.

Jets are reconstructed using the anti-$k_{t}$ algorithm \cite{bib:Cacciari2008}
with a radius parameter of 0.4. The inputs to the jet clustering are built by
combining the information from both the calorimeters and the ID using a
particle-flow algorithm~\cite{PERF-2015-09}.
The cluster energies are calibrated according to in situ measurements of the
jet energy scale~\cite{PERF-2016-04}. Jets with $\pT < 60$~\si{\GeV}
and with $|\eta| <$ 2.4 are further required to satisfy the jet vertex tagger~\cite{PERF-2014-03},
which is a likelihood discriminant that uses a combination of track and vertex
information to suppress jets originating from pile-up activity.

Hadronic decays of $\tau$-leptons are composed of a neutrino and a set of visible decay
products ($\tau_{\text{had}}$), typically one or three charged pions and up
to two neutral pions. The reconstruction of $\tau_{\text{had}}$ starts with jets reconstructed from topological
clusters~\cite{bib:TauReconstructionRun2,ATLAS-CONF-2017-029}.
The $\tau_\mathrm{had}$ candidates must have $|\eta|<2.5$ with the
transition region ($1.37<|\eta|<1.52$) excluded, a transverse momentum $p_{\mathrm{T}} > 65$~\si{\GeV}, one or three associated tracks with $\pm$1 total electric charge.
To discriminate against jets, $\tau_{\text{had}}$ candidates are identified with a  multivariate
algorithm that employs a Recurrent Neural Network (RNN) using shower shape and
tracking information~\cite{ATLAS-CONF-2017-029}.
All $\tau_\mathrm{had}$  candidates are required to satisfy the \emph{Medium}
operating point which corresponds to an efficiency of 75\% (60\%) for 1-prong (3-prong)
$\tau$-leptons, and a jet background rejection of 35 (240) for 1-prong (3-prong) $\tau$-lepton candidates.
Furthermore, a dedicated Boosted Decision Tree-based veto is applied to reduce the number
of electrons misidentified as $\tau_{\text{had}}$ candidates.

Jets with $|\eta|<2.5$ containing a $b$-hadrons are identified with a $b$-tagging
algorithm based on a deep-learning neural network \cite{ATLAS-CONF-2014-046}.
The  information used includes distinctive features of $b$-hadron decays in terms of
the impact parameters of the tracks and the displaced vertices reconstructed in the
ID. The chosen operating point has an efficiency of 85\%. In this study,
a veto on $b$-tagged jets is applied to reject contributions from events containing top quarks.

The missing transverse momentum vector $\vec{p}_{\mathrm{T}}^{\mathrm{miss}}$,
with magnitude $E_{\mathrm{T}}^{\mathrm{miss}}$, is defined as the
negative vector sum of the transverse momenta of all identified physics
objects and an additional soft term. The soft term is constructed from all
tracks that are associated with the primary vertex, but not associated with any
selected physics object~\cite{ATLAS-CONF-2018-023}.

Candidate signal events must have a reconstructed primary vertex with at least two associated tracks, defined as the
vertex whose constituent tracks have the highest sum of $p_{\mathrm{T}}^2$.
There should be exactly two different-flavour, opposite-sign leptons satisfying
the previously mentioned criteria: $e\mu$, $e\tau_{\text{had}}$ or $\mu\tau_{\text{had}}$.
Events with an additional electron, muon or $\tau_\mathrm{had}$ fulfilling all the selections are vetoed.
For the $e\mu$ channel only, events with an extra electron or muon fulfilling the
``loose'' criteria are also vetoed.
For all three channels, the lepton candidates must be back-to-back in
the transverse plane with $\Delta\phi(\ell,\ell^{\prime})>2.7$. The invariant
mass of the dilepton pair larger than 600 \GeV is used as the discriminant.
To account for differences between data and simulation, corrections are applied
to the lepton trigger, reconstruction, identification, and isolation
efficiencies as well as to the lepton energy/momentum resolution and scale
\cite{PERF-2015-10,ATLAS-CONF-2014-032,bib:TauReconstructionRun2}.

To avoid double counting among objects, a lepton-lepton and lepton-jet overlap removal
is applied based on the $\Delta R $ distance metric. The $p_{\mathrm{T}}$ threshold of
muons considered in the overlap removal with $\tau_{\text{had}}$ is lowered to 2 \GeV. Further details can be found in Ref.~\cite{HIGG-2019-09}.

After the event selection mentioned above, for a \Zprime boson with a mass of \SI{1.5}{\TeV}, the fractions of signal events that
satisfy all of the selection requirements are approximately 45\%, 20\%, and
15\% for the $e\mu$, $e\tau$, and $\mu\tau$ final states, respectively.

Because of the presence of a neutrino in the $\tau$-leptons hadronic decay, for the $e\tau$ and $\mu\tau$ channels
the dilepton invariant mass cannot be fully reconstructed, so an approximation is used.
The mass of the BSM particle from the signals considered is much larger than the masses of the SM leptons,
thus producing very high energy $\tau$-leptons when it decays.
The hadronic cascade decay of the high energy $\tau$-lepton results in
the neutrino and the resultant jet being nearly collinear. The neutrino
four-momentum is reconstructed from the magnitude of the missing transverse
momentum and the direction of the 
$\tau_{\text{had}}$ candidate. This
technique improves the dilepton mass resolution and search
sensitivity \cite{SUSY-2013-06}. For a simulated \Zprime boson with a mass of
\SI{2}{\TeV}, the mass resolution improves from  8\% (17\%) to 4\% (12\%) in
the $e\tau$ ($\mu\tau$) channel.


\section{Background estimation}
\label{sec:background}

The background processes for this search can be divided into two categories:
reducible and irreducible. The latter is composed of processes, which
can produce two different-flavour opposite-sign prompt leptons, including
diboson, \ttbar, single-top \(tW\), and $\Z/\gamma^* \to\tau\tau$
production. These processes are modelled with MC simulated event samples
and normalised to their theoretical cross-sections before further adjustment. Reducible
backgrounds originate from jets mis-reconstructed as leptons, such as $W$+jets
and multijet production. These backgrounds are estimated from the data. The contribution from reducible backgrounds
is small in the $e\mu$ channel---approximately 10\% in the signal region (SR)---whereas in the $e\tau$ and $\mu\tau$
channels they are among  the leading components and account for  20$-$30\% of the
total background in the SR. Table~\ref{tab:FitCRSR} shows the regions used in the final fit mentioned in Section~\ref{sec:statistic} in different channels.

\begin{table}[htb]
\begin{center}
\caption[]{Regions used in the final fit mentioned in Section~\ref{sec:statistic} in $e\mu$, $e\tau$ and $\mu\tau$ channels. ``SR'' and ``CR'' refer to signal and control regions respectively.}
\label{tab:FitCRSR}
\begin{tabular}{p{3cm}|p{3cm}|p{8cm}}
\toprule
Region & Channels & Requirements \\
\midrule
\multicolumn{3}{c}{ Nominal $\Delta\phi_{\ell\ell'}$ } \\
\midrule
SR & $e\mu$, $e\tau$ and $\mu\tau$ & $\Delta\phi_{\ell\ell'} > 2.7$, no $b$-jet, $m_{\ell\ell'}$ > 600 \GeV \\
$t\bar{t}$ CR & $e\mu$, $e\tau$ and $\mu\tau$ & $\Delta\phi_{\ell\ell'} > 2.7$, at least one $b$-jet, $m_{\ell\ell'}$ > 600 \GeV  \\
\midrule
\multicolumn{3}{c}{ Reversed $\Delta\phi_{\ell\ell'}$ } \\
\midrule
Low $\Delta\phi_{\ell\ell'}$ $t\bar{t}$ CR & $e\mu$ & $\Delta\phi_{\ell\ell'} < 2.7$, at least one $b$-jet, $m_{\ell\ell'}$ > 600 \GeV   \\
$WW$ CR & $e\mu$ & $\Delta\phi_{\ell\ell'} < 2.7$, no $b$-jet, $m_{\ell\ell'}$ > 600 \GeV \\
\bottomrule
\end{tabular}
\end{center}
\end{table}

\subsection{Irreducible backgrounds}
Among the irreducible backgrounds, \ttbar and $WW$ are the dominant processes in all channels. Their contributions are
evaluted with MC simulation, which are corrected by factors derived in the relevant control regions (CR). For the \ttbar process, which is the dominant contribution in ``Top Quarks'' backgrounds,
a CR is built with the same
selection as signal events but reversing the $b$-jet veto cut in each of
$e\mu$, $e\tau$ and $\mu\tau$ channels. 
In the $e\mu$ channel, a $WW$ CR is built with the SR selection, but reversing
the $\Delta\phi_{\ell\ell'}$ cut. In this $WW$ CR, as the \ttbar contamination is significant ($\sim$ 40\%),
a dedicated \ttbar CR is defined in the $\Delta\phi_{\ell\ell'}<2.7$ region to control the \ttbar contamination.
For the $e\tau$ and $\mu\tau$ channels, $WW$ CRs are not defined
due to non-negligible contributions from fake backgrounds. However, due to lepton flavour universality,
the correction factor ratios between the \ttbar and $WW$ processes are assumed to be the same in the $e\mu$ and $\tau$-lepton channels:
\begin{equation}
\frac{k_{WW}^{e(\mu)\tau}}{k_{WW}^{e\mu}}=\frac{k_{t\bar{t}}^{e(\mu)\tau}}{k_{t\bar{t}}^{e\mu}},
\end{equation}
where $k_{WW}^{e(\mu)\tau}$ is the correction factor for the $WW$ process in the $e\tau$ or $\mu\tau$ channel, $k_{WW}^{e\mu}$ is the correction factor for the $WW$ process in the $e\mu$ channel. The $k_{t\bar{t}}^{e(\mu)\tau}$ is the correction factor for the \ttbar process in the $e\tau$ or $\mu\tau$ channel, and $k_{t\bar{t}}^{e\mu}$ is the correction factor for the \ttbar process in the $e\mu$ channel.
The correction factor for the $WW$ process in the $e\mu$ channel can be extrapolated to the $e\tau$ or $\mu\tau$  channel with a factor equal to the ratio of the \ttbar correction factors between the $e\mu$ channel and $\tau$ channels:
\begin{equation}
k_{WW}^{e(\mu)\tau}=k_{WW}^{e\mu}\times\frac{k_{t\bar{t}}^{e(\mu)\tau}}{k_{t\bar{t}}^{e\mu}}.
\end{equation}
From the simultaneous fit in all CRs for the $e\mu$ channel, the fitted uncertainty of
$k_{WW}^{e\mu}$ is 12\% and it is highly correlated with $k_{t\bar{t}}^{e\mu}$.
The same uncertainty is assigned to $k_{WW}^{e(\mu)\tau}$. Conservatively, it is treated
as uncorrelated to $k_{t\bar{t}}^{e\mu}$, thus leading to an over-estimate in the final fit uncertainties.
The 12\% uncertainty on the $WW$ correction is also much larger than the
correction effect itself in the $e\tau$ and $\mu\tau$ channels, which is less than 5\% on the diboson background.
 
The \ttbar and $WW$ correction factors are used to correct the total yields of ``Top Quarks'' and
``Diboson'' backgrounds respectively.
Other irreducible background processes, such as $Z\to\ell\ell'$ are evaluted directly with MC simulation.

\subsection{Reducible backgrounds}
 
The dominant reducible backgrounds, \W+jets and multijet production, are estimated by using data-driven techniques. In the $e\mu$ channel,
the matrix method~\cite{EXOT-2012-23} is used, while in the $e\tau$ and $\mu\tau$ channels, the fake-lepton background is extrapolated from dedicated CRs to SR with the method described in Ref.~\cite{SUSY-2013-06}.
 
\subsubsection{Fake backgrounds in $e\mu$ channel}
 
The goal of the matrix method is to estimate the
fraction of events in the data sample with a ``real'' electron and a ``real'' muon ($N_{\text{RR}}$),
events with a jet faking an electron (``fake'' electron) and a ``fake'' muon ($N_{\text{FF}}$), and events with
one ``real'' lepton and one ``fake'' lepton ($N_{\text{RF}}$ and $N_{\text{FR}}$).
 
Two selection criteria are defined: ``Tight'' and ``Loose'' for muons (electrons) based on their lepton quality (identification and isolation) respectively. As the selection efficiency is different between a ``real'' lepton and a ``fake'' lepton, the contribution of ``fake'' leptons is estimated by the number of data events passing each selection criteria.
 
Four data samples are constructed based on the lepton quality:
\begin{itemize}
\item{a sample where both the electrons and muons pass the ``Tight'' quality cut, with number of events $N_{\text{TT}}$; }
\item{a sample where the electrons pass the ``Loose'' but failed the ``Tight'' selection, and the muons pass
the ``Tight'' selection, with event number $N_{\text{LT}}$;}
\item{a sample where the muons pass the ``Loose'' but failed the ``Tight'' selection, and the electrons pass
the ``Tight'' selection, with event number $N_{\text{TL}}$;}
\item{a sample where both the electrons and muons pass the ``Loose'' but fail the ``Tight'' selection, with
event number $N_{\text{LL}}$. }
\end{itemize}
 
By solving a system of linear equations involving the numbers of events for these four data samples,  the reducible
background contributions can be evaluated.
\begin{gather}
\begin{bmatrix}
$$N_{\mathrm{TT}}$$ \\ $$N_{\mathrm{LT}}$$ \\ $$N_{\mathrm{TL}}$$ \\ $$N_{\mathrm{LL}}$$
\end{bmatrix} =
\begin{bmatrix}
$$r_{e}r_{\mu} & f_{e}r_{\mu} & r_{e}f_{\mu} & f_{e}f_{\mu} \\
(1-r_{e})r_{\mu} & (1-f_{e})r_{\mu} & (1-r_{e})f_{\mu} & (1-f_{e})f_{\mu} \\
r_{e}(1-r_{\mu}) & f_{e}(1-r_{\mu}) & r_{e}(1-f_{\mu}) & f_{e}(1-f_{\mu}) \\
(1-r_{e})(1-r_{\mu}) & (1-f_{e})(1-r_{\mu}) & (1-r_{e})(1-f_{\mu}) & (1-f_{e})(1-f_{\mu}) $$
\end{bmatrix}
\begin{bmatrix}
$$N_{\mathrm{RR}}$$ \\ $$N_{\mathrm{FR}}$$ \\ $$N_{\mathrm{RF}}$$ \\ $$N_{\mathrm{FF}}$$
\end{bmatrix}
\end{gather}\;
 
Among these equations, ``$r$'' is the probability of a ``Loose'' lepton matched to a
true lepton to satisfy the ``Tight'' quality selection, defined as the ``real efficiency.''
It is evaluated from $Z\to \ell\ell'$ simulated events. The ``$f$'' refers to the probability that a jet
is misidentified as a ``Tight'' quality lepton, so called ``fake rate'' which is determined in a
multijet-enriched data sample. The creation of this sample starts with the selection of a
``Loose'' lepton back-to-back with a jet. To suppress the \W+jets contamination to the multijet-enriched
sample, the events are required to have $E_{\mathrm{T}}^{\mathrm{miss}}<25$~\si{\GeV} and the transverse mass $m_{\mathrm{T}}<50$~\si{\GeV}. Transverse mass is defined as $m_{\mathrm{T}}=\sqrt{2E_{\mathrm{T}}^{\mathrm{\ell^{\pm}}}E_{\mathrm{T}}^{\mathrm{miss}}(1-\cos{\Delta{\phi}})}$, where $E_{\mathrm{T}}^{\mathrm{\ell^{\pm}}}$ and $E_{\mathrm{T}}^{\mathrm{miss}}$ denote the transverse energies of the final state charged lepton and neutrino with  $\Delta{\phi}$ as the azimuthal angle difference. Since  most of the ``fake''
muons are from heavy flavour jets decays in the muon fake rate measurement, the requirements on both of the impact parameters $d_0$ and $z_0$ are
reversed in order to further suppress the \W+jets contamination. Remaining contamination are subtracted relying on simulations from \W+jets and other SM background processes (top, diboson and $Z\to\ell\ell$).
Both the real efficiency and the fake rate are determined as a function of $p_{\mathrm{T}}$.
 
The uncertainties associated with the matrix method are evaluated by considering systematic effects on the
lepton fake rate measurement and uncertainties in the real lepton efficiency.
The latter is considered to be the difference in the lepton $\eta$ distribution between SR and $Z\to\ell\ell$
MC \cite{HIGG-2019-09}. It is assigned by comparing the real
lepton efficiency measured with $Z\to\ell\ell$
plus zero jet events and inclusive $Z\to\ell\ell$ events.
The systematic uncertainties in the fake rate include:
\begin{itemize}
\item{the choice of multijet-enriched region;}
\item{the impact of reversing $d_0$ and $z_0$ on muons;}
\item{the difference in the fake rates between the SR and multijet-enriched region due to the difference in the jet flavour components.}
\end{itemize}
The overall uncertainty on the $e\mu$ reducible background is about 30\% in the SR. Given that in the $e\mu$ channel
SR, this contribution is about 10\% of the total background over the invariant mass range considered, the
uncertainties in the fake-lepton background estimates have little impact on the results.
 
\subsubsection{Fake backgrounds in $e\tau$ and $\mu\tau$ channels}
 
One of the major backgrounds for the $e\tau$ and $\mu\tau$ channels is the \W+jets
process, where a jet is misidentified as a lepton candidate. As it is difficult to model misidentification
of jets as leptons, particularly in  the high $p_{\mathrm{T}}$ region, the \W+jets backgrounds are determined
from dedicated CRs. 
They are built from events selected with the same criteria as used in the signal selection except for requiring $E_{\mathrm{T}}^{\mathrm{miss}}>30$~\si{\GeV}
to enhance the \W+jets contribution and suppress the $Z\to\ell\ell'$ contamination. The dilepton mass requirement is also lowered to 130~\si{\GeV} to reduce the statistical uncertainty in these control regions.
Furthermore, the $\Delta\phi_{\ell\ell'}$ requirement is reversed ($\Delta\phi_{\ell\ell'}<2.7$) to ensure
orthogonality with the SR. Simulation studies indicate that the multijet background is negligible
in this CR\@. 
Contributions from other SM background processes are estimated by using MC simulation. The \W+jet MC samples are used to calculate the ratio of events
in the CR to events in the SR\@.
Then the total \W+jet backgrounds yields are extrapolated from CR to SR with this ratio.
Kinematic distributions of the \W+jets background are derived by relying on MC simulation.
The uncertainties considered in this method include
\begin{itemize}
\item the construction of the \W+jets CR,
\item uncertainties in the estimates of other SM background processes in the \W+jets CR,
\item and the \W+jets MC shape uncertainty which is assigned based on the data versus prediction agreement in the \W+jet CR.
\end{itemize}
The overall uncertainty of \W+jets in $e\tau$ and $\mu\tau$ channels is approximately 20\%.
 
The contribution from multijets production is estimated also from a CR with the same selection as the
SR except that the leptons are required to have the same electric charge, assuming
that the probability of misidentifying a jet as a lepton is independent of the charge. The dilepton mass requirement is also lowered to 130~\si{\GeV} to estimate the low mass multijets contribution used in the high mass extrapolation.
The contamination from processes with prompt leptons is subtracted from data in the same-sign CR
using MC simulation, while the contamination from \W+jets is subtracted using the previously mentioned procedure.
The assumption of charge independence
of the jet misidentification rates is tested in a multijet-enriched region with two non-isolated leptons.
After subtraction of other SM backgrounds, the ratio of opposite-sign to same-sign events is found to be
in reasonable agreement in the range of 10\% to 20\% up to the \TeV dilepton invariant mass range.
The remaining lack of closure is taken as an uncertainty.
Besides this charge independence assumption uncertainty, the uncertainties in multijets estimates mainly
come from uncertainties in the \W+jets subtraction and other MC irreducible background subtractions.
As the uncertainty in \W+jets is quite significant, they are also propagated to the multijets estimate
through the \W+jets subtraction.  In this way, the yield of the multijets background is highly anti-correlated with
the \W+jets background, and they are combined together as ``fake'' background. The overall uncertainty in the ``fake'' background is approximately 25\% in the $e\tau$ and $\mu\tau$ channels.

Considering the lack of ``fake'' background event numbers at high dilepton invariant mass values in the signal region, a function of the form $f(m_{l\tau})=Am_{l\tau}^{B+C\mathrm{log}(m_{l\tau})}$ with free parameters $A$, $B$ and $C$ is fitted to the dilepton invariant mass distribution in the range $340 < m_{l\tau} < 3000$ \GeV. This function is used to extrapolate the fake background estimate for the whole signal region ($m_{l\tau}>600$ \GeV). The statistical uncertainty of the fitted function is determined by the propagation of the statistical uncertainties on the fitted parameters. The uncertainty due to the extrapolation is dominated by the fit function choice uncertainty, and it is evaluated by comparing the nominal estimate with the one obtained when changing the fit function ($f(m_{l\tau})=Am_{l\tau}^{B}$, with the free parameters $A$ and $B$). The difference of these two functions is more than 100\% above 2 \TeV. The same functions have been previously used to model the jet backgrounds in other analyses~\cite{EXOT-2017-06}. Both the data-driven uncertainties and the fit related uncertainties are taken into account as independent sources for the fake estimation uncertainty.


\section{Systematic uncertainties}

Systematic uncertainties affect the event yields and the shape of the invariant mass distribution in the signal and control regions. They are grouped into two types: theoretical and experimental uncertainties. The systematic uncertainties related to the estimates of fake backgrounds are discussed in Section~\ref{sec:background}.

Experimental uncertainties considered include those from the trigger, reconstruction, identification and isolation requirements of the final-state particles, such as electrons~\cite{ATLAS-CONF-2014-032}, muons~\cite{PERF-2015-10}, $\tau$-leptons~\cite{bib:TauReconstructionRun2}, jets~\cite{PERF-2016-04}, $E^{\mathrm{miss}}_{\mathrm{{T}}}$~\cite{ATL-PHYS-PUB-2015-027}, and b-tagging~\cite{ATL-PHYS-PUB-2015-022}. Their energy scale and resolution uncertainties are also taken into account. An additional uncertainty (1.7\%) from the measurement of the integrated luminosity 
is included. Sources of uncertainty are considered for both the simulated background and signal processes. Experimental uncertainties affect the event yields of background and the signal cross-section through their effects on the acceptance and the event migration between different analysis regions.

Theoretical uncertainties are considered for the renormalisation ($\mu_r$) and factorisation ($\mu_f$) scales, the choice of PDF, the choice of the value of the strong coupling constant $\alpha_s$, and the modelling of the $t\bar{t}$ and diboson backgrounds. The strategy used in the same analysis with partial data sample at $\sqrt{s}=13$ TeV~\cite{EXOT-2016-09} is followed in this search.
 
Variation of the renormalisation and factorisation scales is used to estimate the uncertainty due to missing higher order corrections. Pairwise variations of $\mu_r$ and $\mu_f$ are considered to find the maximum and minimum variations. The PDF uncertainty consists of the contribution from the PDF set used in the matrix element calculation. It is estimated using different PDF sets and eigenvector variations within a particular PDF set for the top-quark, diboson, and \W+jets backgrounds. The $\alpha_s$ uncertainty is evaluated by using the same PDF set evaluated with two different $\alpha_s$ values. The uncertainty in the modelling of the \ttbar background is assessed by evaluating two MC samples which are generated by \powhegbox+\pythia8 and \AMCatNLO+\pythia8 as sources of modelling uncertainty. The uncertainty in the modelling of the diboson background is assessed by evaluating two MC samples which are generated by \textsc{Sherpa} 2.2.12 and \powhegbox+\pythia8.

Experimental systematic uncertainties common to signal and background processes are assumed to be correlated. For signal processes, only experimental systematic uncertainties and uncertainties due to the limited statistical precision of simulated samples are considered. All uncertainties are evaluated as a function of $m_{\ell\ell^{\prime}}$.
 
For illustration, Table~\ref{tab:syst_sum} lists the post-fit impact of uncertainties in the RPV SUSY $\tilde{\nu}_{\tau}$ model for each measurement grouped by their respective sources. The \Zprime and QBH models have similar results. The sources of the largest systematic uncertainties 
are the statistical uncertainties associated with the background estimate and background modelling uncertainties in the $e\mu$ channel. In the $\tau$-lepton channels, the uncertainties are dominated by the statistical precision of the background estimate and experimental uncertainties related to $\tau$-leptons and the estimate of fake backgrounds. The total uncertainties are dominated by the statistical precision in all three channels.

\begin{table}[htbp]
\begin{center}
\caption{Summary of the different sources of relative uncertainty (post-fit) in percentage on the observed signal strength of the RPV SUSY $\tilde{\nu}_{\tau}$ model with a mass of \SI{3}{\TeV} in the $e\mu$, $e\tau$, and $\mu\tau$ channels. ``Other'' refers to luminosity, jet-vertex-tagger (JVT) and pile-up weight uncertainties. ``NA'' stands for not applicable.}
\label{tab:syst_sum}
\resizebox{\textwidth}{!}{
\begin{tabular}{l|c|c|c}
\toprule
Source of uncertainty (in percent)    & Impact on observed $\mu_{\mathrm{RPV}}^{e\mu}$ & Impact on observed $\mu_{\mathrm{RPV}}^{e\tau}$ & Impact on observed $\mu_{\mathrm{RPV}}^{\mu\tau}$ \\
\midrule
Electrons                                & 2.2  & 0.85  & NA   \\
Muons                                    & 2.8  & NA     & 4.4 \\
$\tau$-leptons                           & NA    & 9.7   & 11   \\
Jets and $E_{\mathrm{T}}^{\mathrm{miss}}$ & 2.1  & 0.8  & 0.8 \\
Flavour tagging                           & 2.1  & <0.1  & 0.1 \\
Other (luminosity, JVT, pile-up)         & 0.6 & 0.4  & 0.7 \\
Fake backgrounds                & 0.6 & 3.2   & 9.7  \\
Background modelling                     & 9.6  & 2.1   & 7.3 \\
Top and Diboson normalisations               & 8.7  & 1.6   & 1.8  \\
Simulation statistical uncertainty                    & 28   & 9.6   & 15  \\
\midrule
Total systematic uncertainty             & 32   & 14   & 23 \\
Data statistical uncertainty                          & 53   & 48   & 71 \\
Total uncertainty                                   & 62   & 50   & 74 \\
\bottomrule
\end{tabular}}
\end{center}
\end{table}


\section{Statistical analysis}
\label{sec:statistic}

The dilepton invariant mass distribution in the SR and CRs are fitted simultaneously to test for the presence of a signal. For hypothesis testing, binned profile-likelihood fits are performed, following a modified frequentist method~\cite{Read:2002hq} implemented in \textsc{RooStats}~\cite{RooFit}.
The fits are performed for each decay channel separately.
For the $e\mu$ channel, $WW$ and low $\Delta\phi_{\ell\ell'}$ \ttbar CRs are included in the fit to extract
the overall normalisation of the diboson background while keeping the normalisation of the top quark background uncorrelated in the high- and low-$\Delta\phi_{\ell\ell'}$ regions. Separate normalisation factors of the top quark background are used for the high- and low-$\Delta\phi_{\ell\ell'}$ regions in the simultaneous fit, by which their correlation is determined. For the $e\tau$ and $\mu\tau$ channels, only $t\bar{t}$ CRs are included in the fit and the diboson correction factors are calculated from the CR-only fit results and put into the final fit as fixed normalisation factors.
 
The binned likelihood function ($\mathcal{L}(\mu,\theta)$) is constructed as a product of Poisson probability terms over all bins considered in the search. The likelihood function depends on the parameter of interest (POI), the signal strength parameter $\mu$, a factor multiplying the theoretical signal production cross-section and a set of nuisance parameters $\theta$ that encode the effect of systematic uncertainties in the signal and background expectations. All nuisance parameters are implemented in the likelihood function as Gaussian constraints. Unconstrained normalisation factors are applied on the top quark and diboson background components in the $e\mu$ channel, and the top quark background in the $\tau$-lepton channels. They are controlled by the previously mentioned CRs in the respective channels. The expected event yield in a bin depends on the normalisation factors and on the nuisance parameters. The nuisance parameters adjust the expected event yields for signal and background according to the best fit to data.
 
The likelihood function is fitted to the data to test for the presence of a signal. The test statistic $q_{\mu}$ is defined as the profile likelihood ratio, $q_{\mu}=-2\ln\mathcal{L}(\mu,\hat{\hat \theta})/\mathcal{L}(\hat{\mu},\hat{\theta})$, where $\hat{\mu}$ and $\hat{\theta}$ are the values of the parameters that simultaneously maximise the likelihood function, and $\hat{\hat \theta}$ are the values of the nuisance parameters that maximise the likelihood function for a fixed value of $\mu$.  Compatibility of the observed data with the background-only hypothesis is tested by setting $\mu = 0$ in the test statistic $q_0$. Upper limits on the signal production cross-section for each considered signal scenario are computed using $q_{\mu}$ in the $\mathrm{CL}_s$ method~\cite{Read:2002hq} with the asymptotic approximation~\cite{Cowan:2010js}. A given signal scenario is considered to be excluded at the 95\% confidence level (CL) if the value of the signal production cross-section (parameterised by $\mu$) yields a $\mathrm{CL}_s$ value less than 0.05.


\section{Results}

Tables~\ref{Tab:CR_emu}, ~\ref{Tab:CR_tau} and ~\ref{Tab:SR_emu_etau_mutau} show the observed data event yields and the expected background event yields in the CRs and SRs for $e\mu$, $e\tau$ and $\mu\tau$ channels.
The $e\mu$ background is dominated by $t\bar{t}$ and diboson events, while \W+jets events are dominant for the $e\tau$ and $\mu\tau$ final states. The \W+jets background in the $e\tau$ and $\mu\tau$ channels is highly suppressed by the RNN $\tau$ identification comparing to the previous analysis.
 
\begin{figure}[!hbtp]
\centering
\subfloat[]{
\includegraphics[width=0.49\columnwidth]{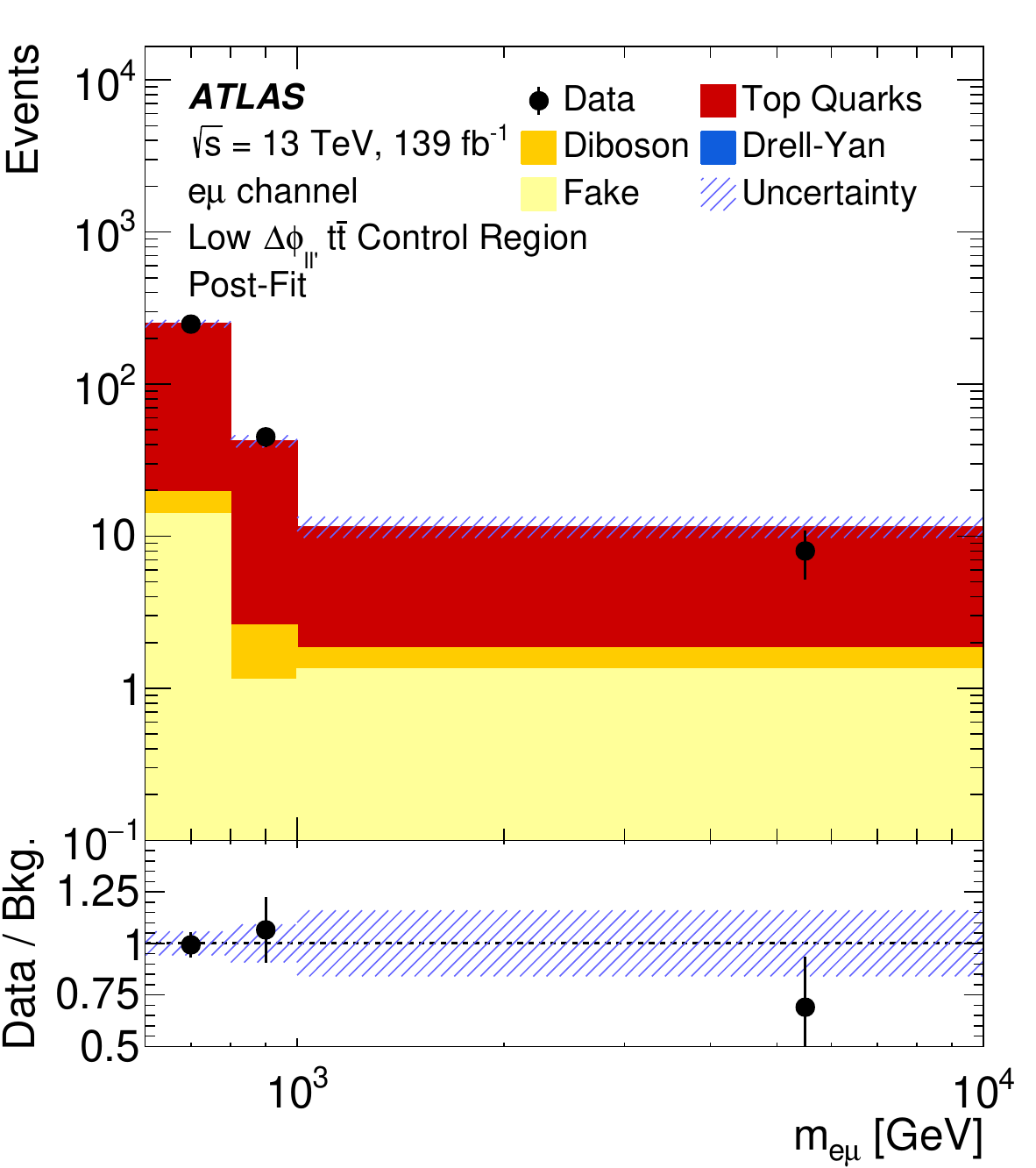}}
\subfloat[]{
\includegraphics[width=0.49\columnwidth]{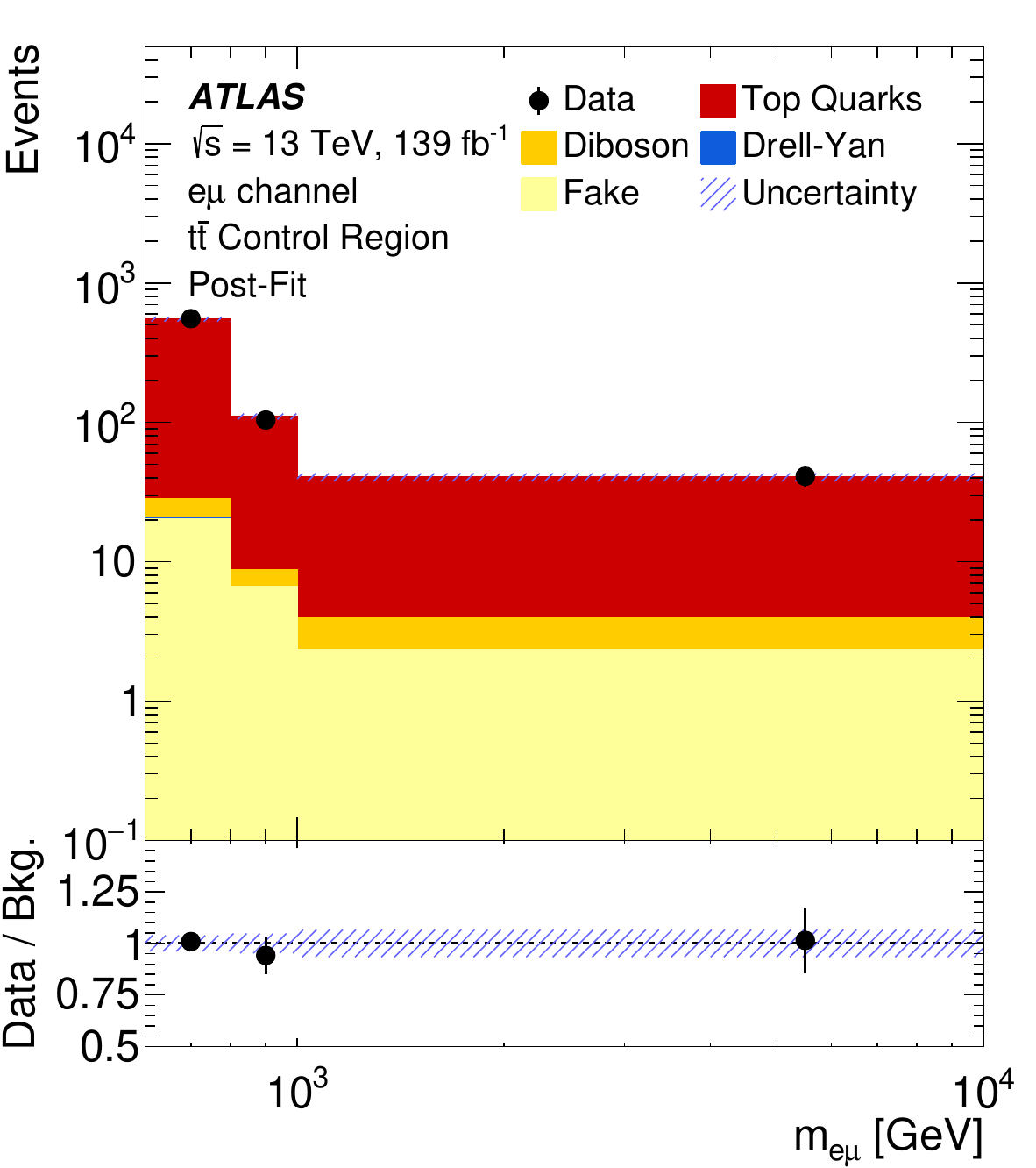}} \\
\subfloat[]{
\includegraphics[width=0.49\columnwidth]{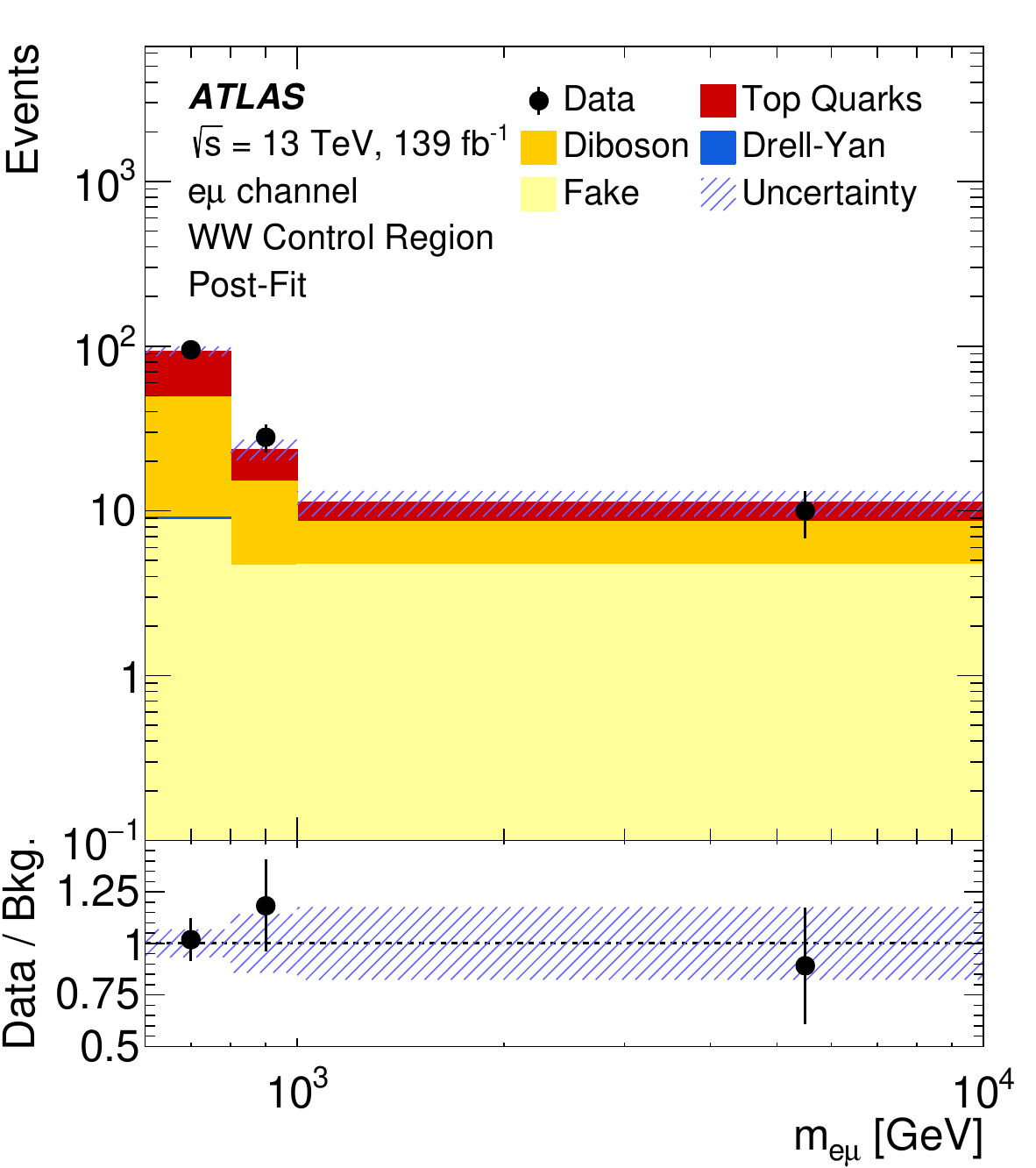}}
\caption{The background-only post-fit invariant mass distribution of (a) Low $\Delta\phi_{\ell\ell'}$ \ttbar CR, (b) \ttbar CR and (c) $WW$ CR for data and the SM background predictions in the $e\mu$ channel. The error bars show the statistical uncertainty of the observed yields, while the hashed band includes the post-fit total uncertainties taking into account all the correlations. The ratio of data to the best-fit prediction is shown in the bottom panels of the plots. The last bin contains the overflow events. The binning of the control regions was limited by the background statistics.}
\label{fig:CRfits_emu}
\end{figure}

\begin{figure}[!hbtp]
\centering
\subfloat[]{
\includegraphics[width=0.49\columnwidth]{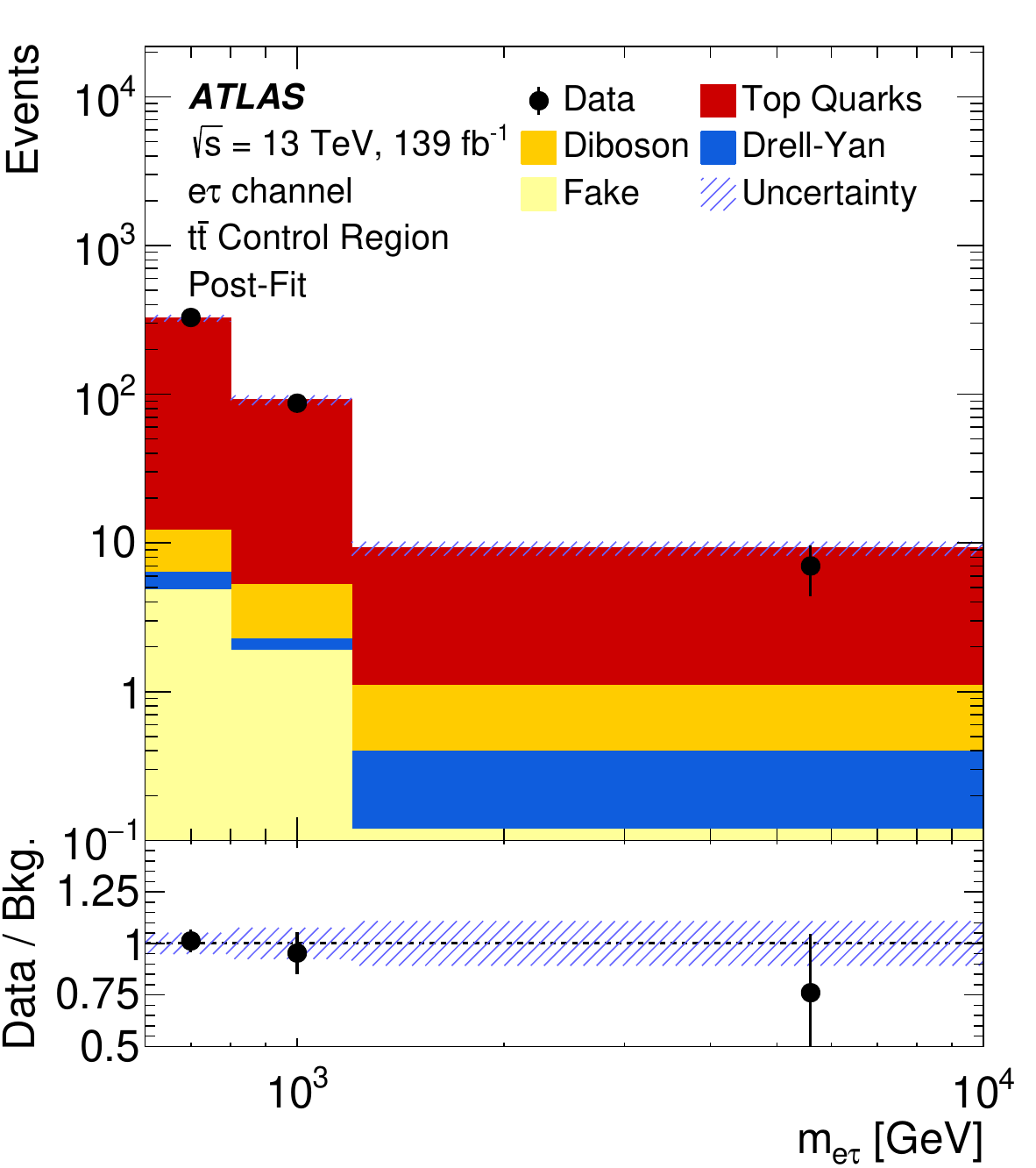}}
\subfloat[]{
\includegraphics[width=0.49\columnwidth]{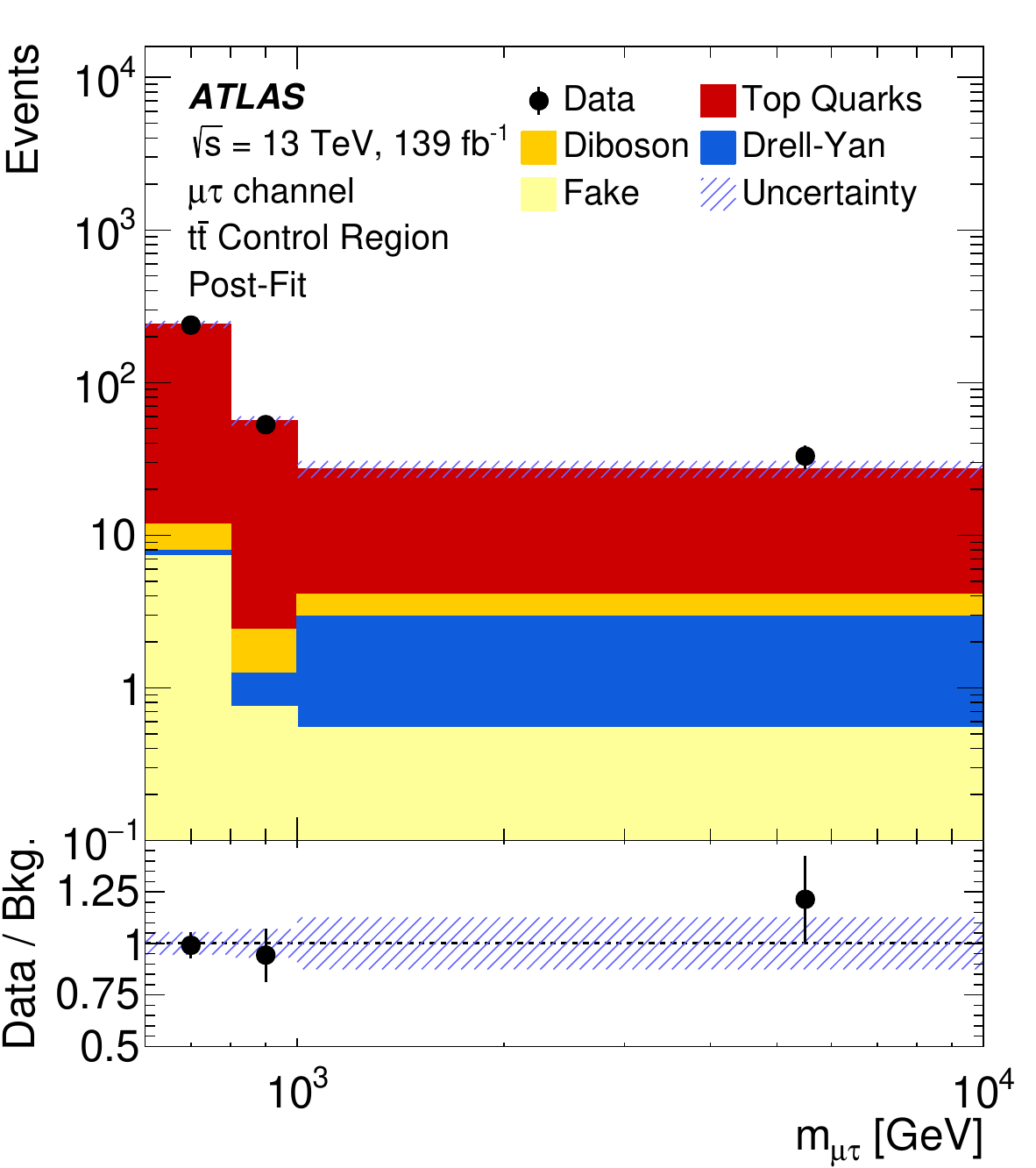}}
\caption{The background-only post-fit invariant mass distribution of (a) $e\tau$ and (b) $\mu\tau$ pairs for data and the SM background predictions in the \ttbar CR. The error bars show the statistical uncertainty of the observed yields, while the hashed band includes the post-fit total uncertainties taking into account all the correlations. The ratio of data to the best-fit prediction is shown in the bottom panels of the plots. The last bin contains the overflow events.}
\label{fig:CRfits_tau}
\end{figure}

\begin{figure}[!hbtp]
\centering
\subfloat[]{
\includegraphics[width=0.49\columnwidth]{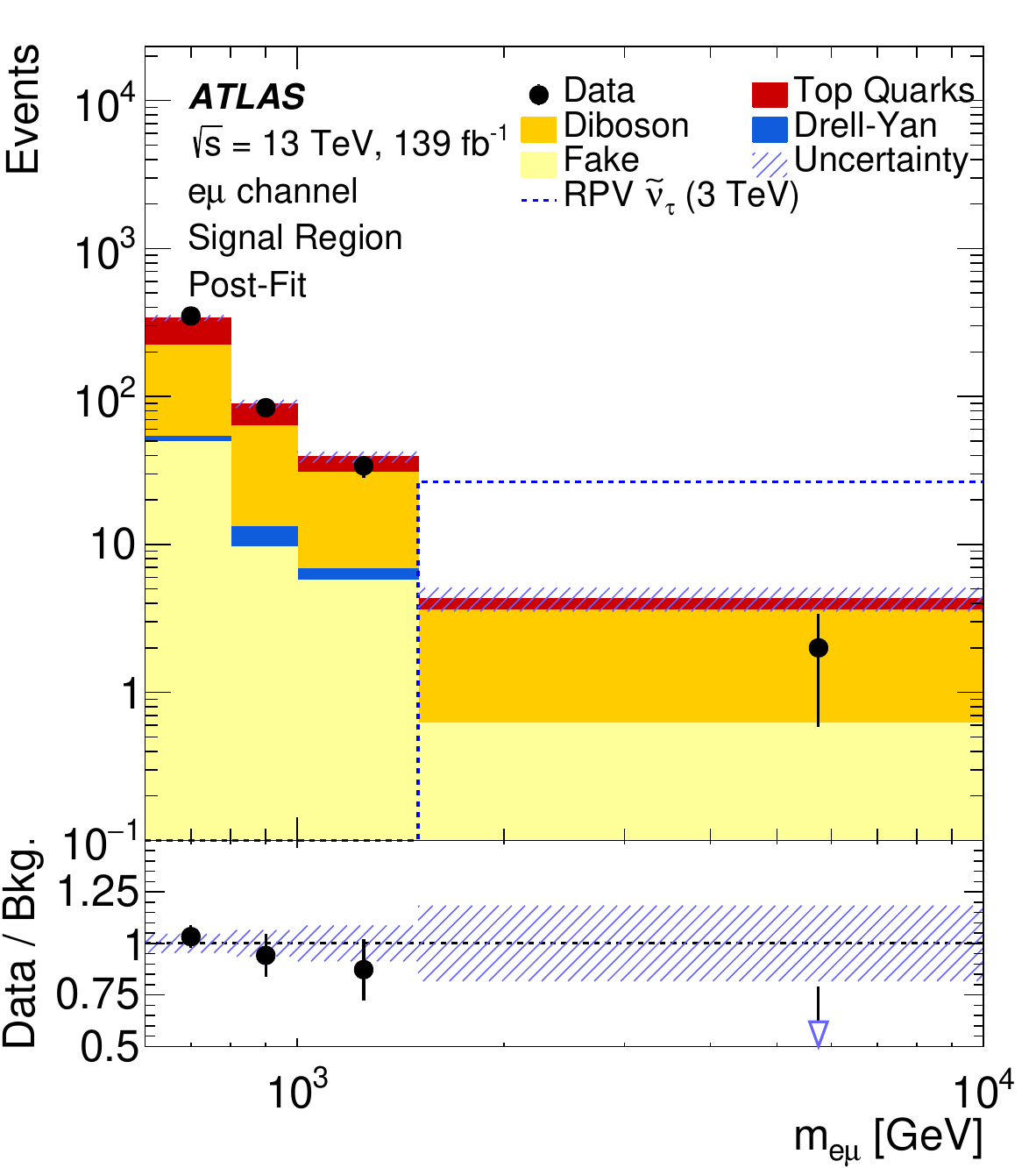}}
\subfloat[]{
\includegraphics[width=0.49\columnwidth]{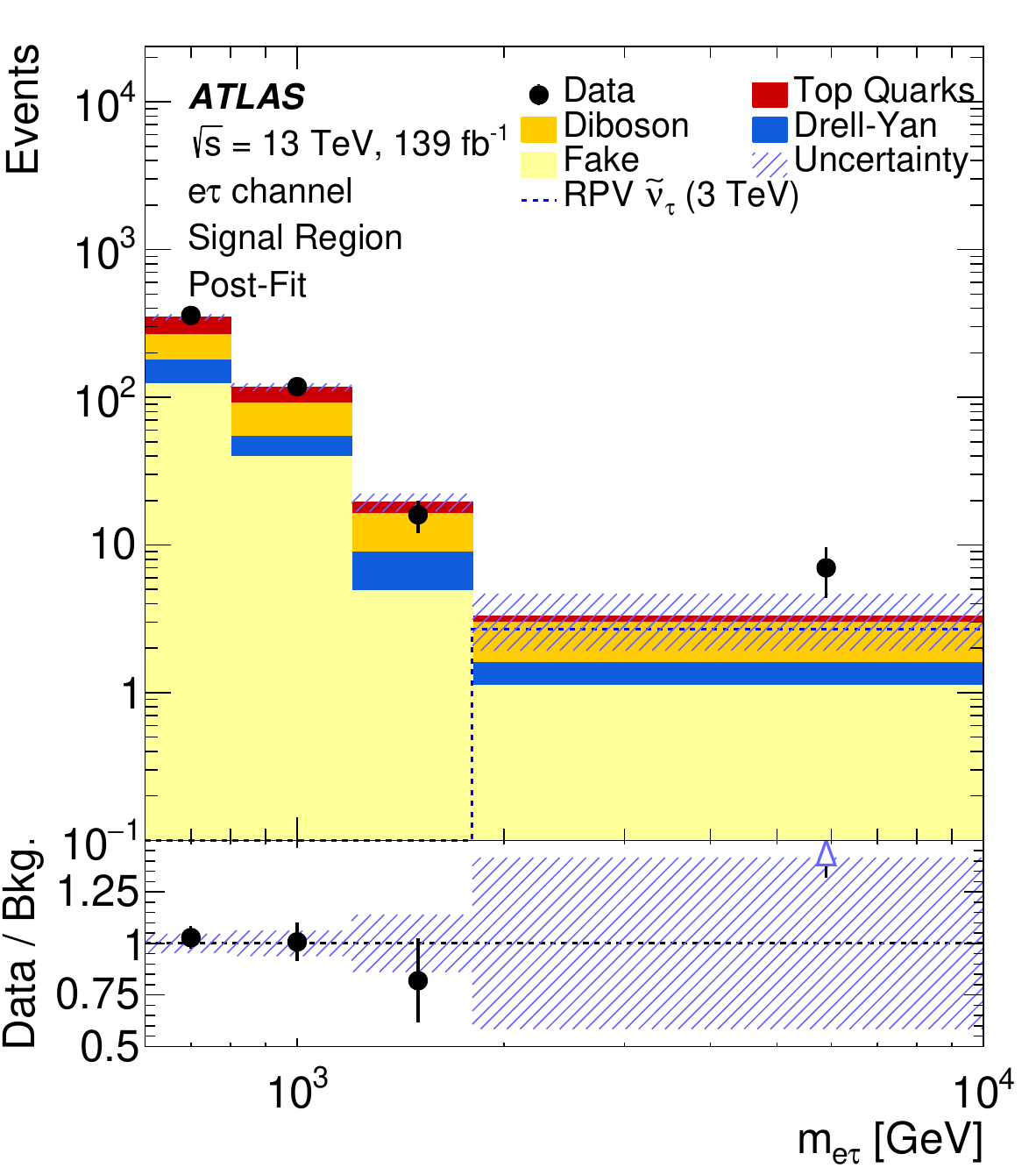}} \\
\subfloat[]{
\includegraphics[width=0.49\columnwidth]{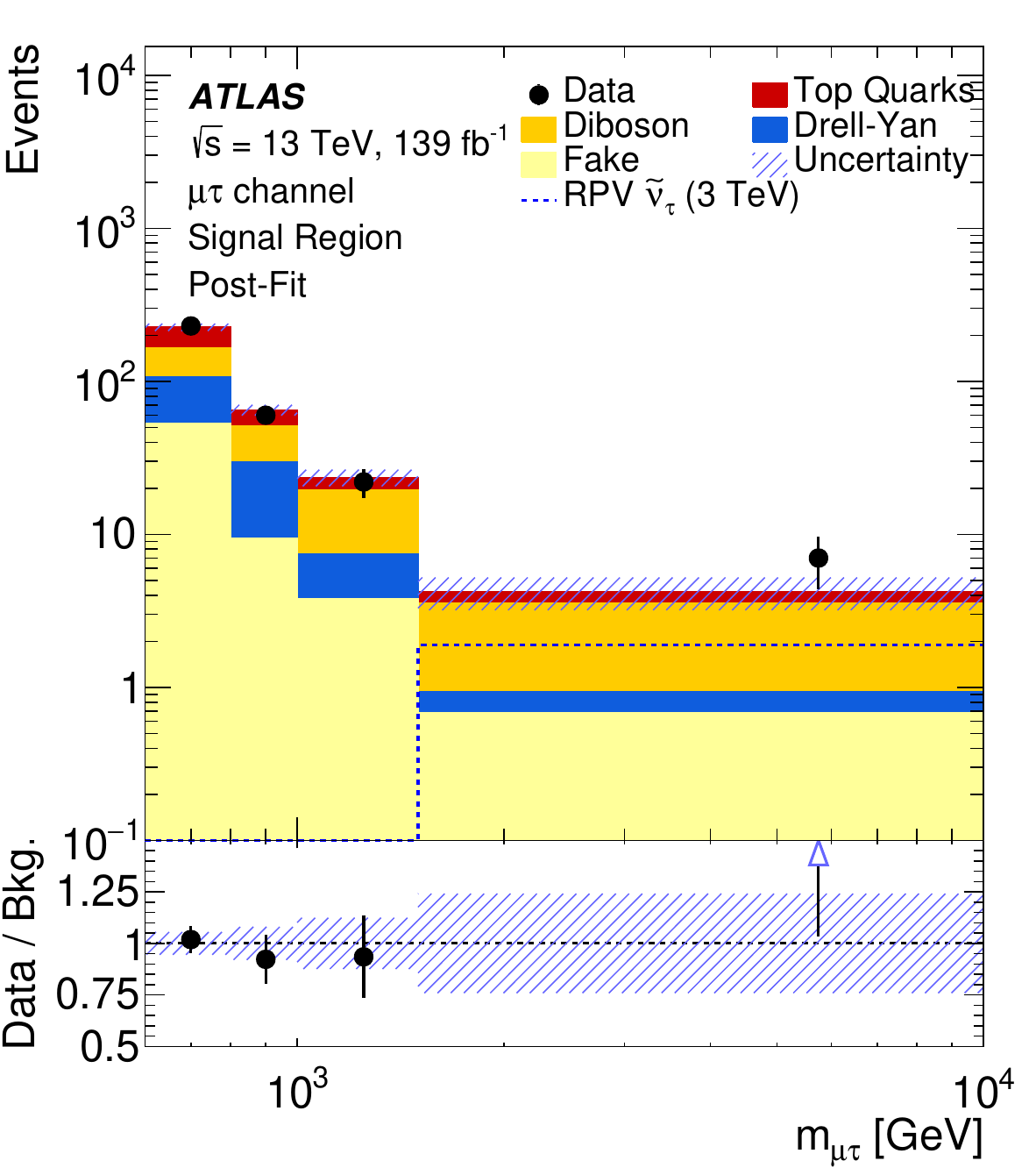}}
\caption{The background-only post-fit invariant mass distribution of (a) $e\mu$, (b) $e\tau$ and (c) $\mu\tau$ pairs for data and the SM background predictions in the SR. The error bars show the statistical uncertainty of the observed yields, while the hashed band includes the post-fit total uncertainties taking into account all the correlations. The dashed line shows a typical pre-fit signal mass distribution (RPV $\tilde{\nu}_{\tau}$ at 3 \TeV). The ratio of data to the best-fit prediction is shown in the bottom panels of the plots. The arrow shows the difference between data and MC which exceed the ratio range. The last bin contains the overflow events.}
\label{fig:invmass}
\end{figure}

Binned profile-likelihood fits are performed on dilepton invariant mass spectra for each signal scenario
in $e\mu$, $e\tau$ and $\mu\tau$ channels. Table~\ref{Tab:NormF} shows the fitted correction factors for ``Top Quarks'' and
``Diboson'' backgrounds applied to SR\@. Figures~\ref{fig:CRfits_emu} and ~\ref{fig:CRfits_tau} show the post-fit dilepton invariant mass distributions of the background-only fit in the CRs for the $e\mu$, $e\tau$ and $\mu\tau$ channels. Figure~\ref{fig:invmass} shows the post-fit dilepton invariant mass distributions of the background-only fit in the SR for the $e\mu$, $e\tau$ and $\mu\tau$ channels. The SM expectations agree well with the data in the SR of the $e\mu$ channel\@. Mild tension with the SM background 
estimates is observed between 2.0 \si{\TeV} and 2.3 \TeV in the SRs of the $\tau$-lepton channels. Nevertheless,  SM expectations are still statistically consistent with the data within 2$\sigma$.
 
\begin{figure}[!htbp]
\centering
\subfloat[]{
\includegraphics[width=0.49\linewidth]{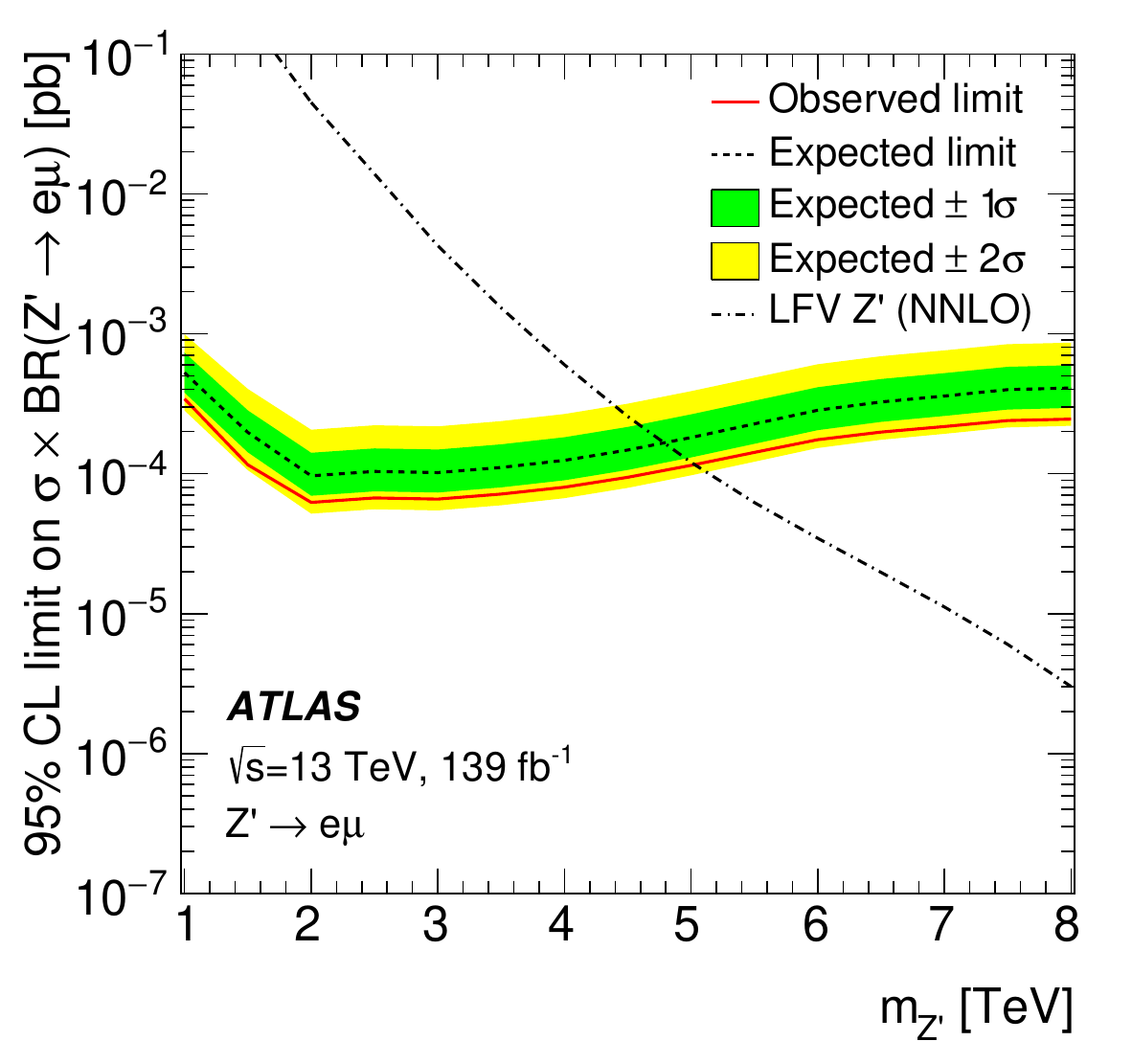}}
\subfloat[]{
\includegraphics[width=0.49\linewidth]{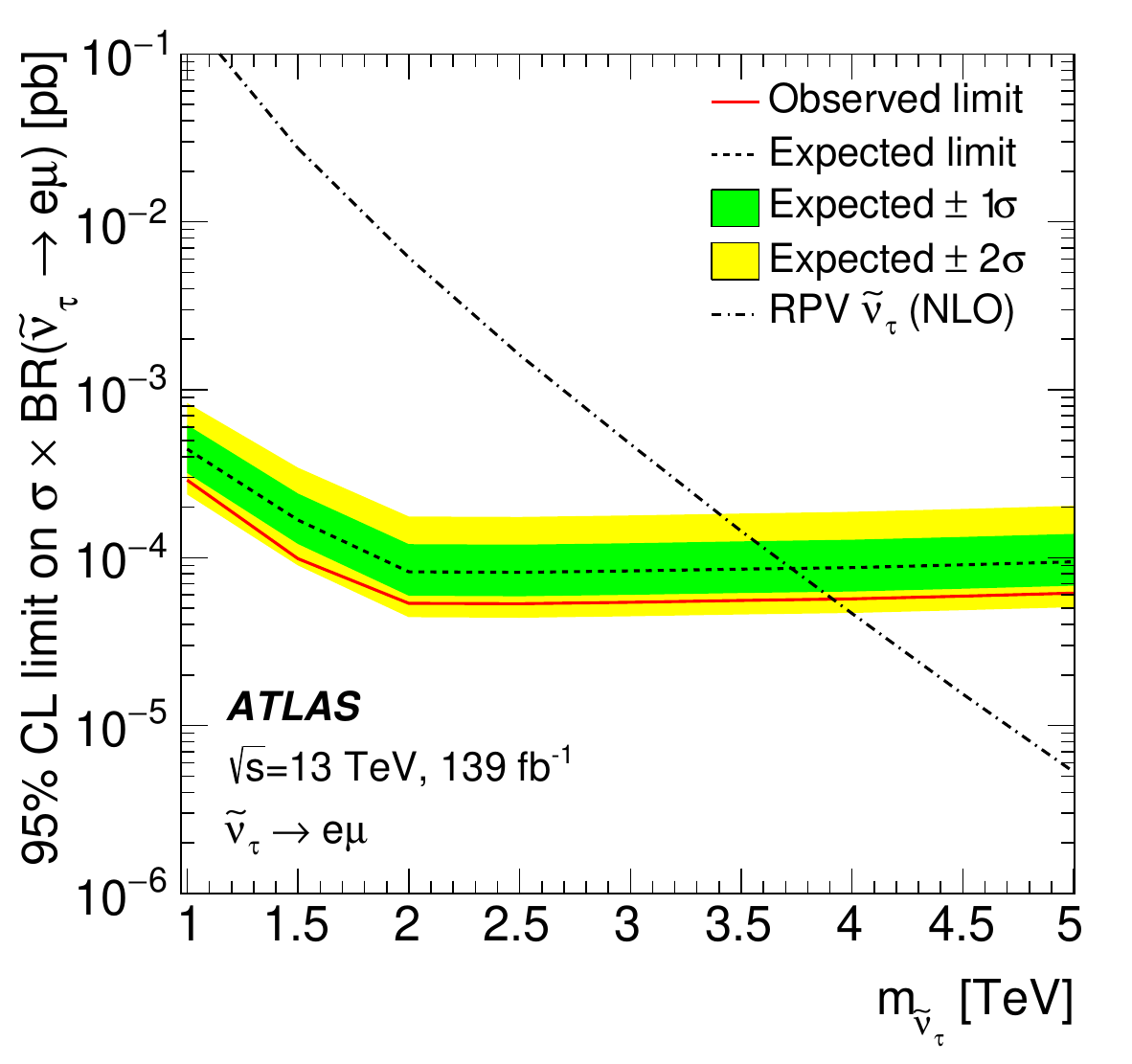}} \\
\subfloat[]{
\includegraphics[width=0.49\linewidth]{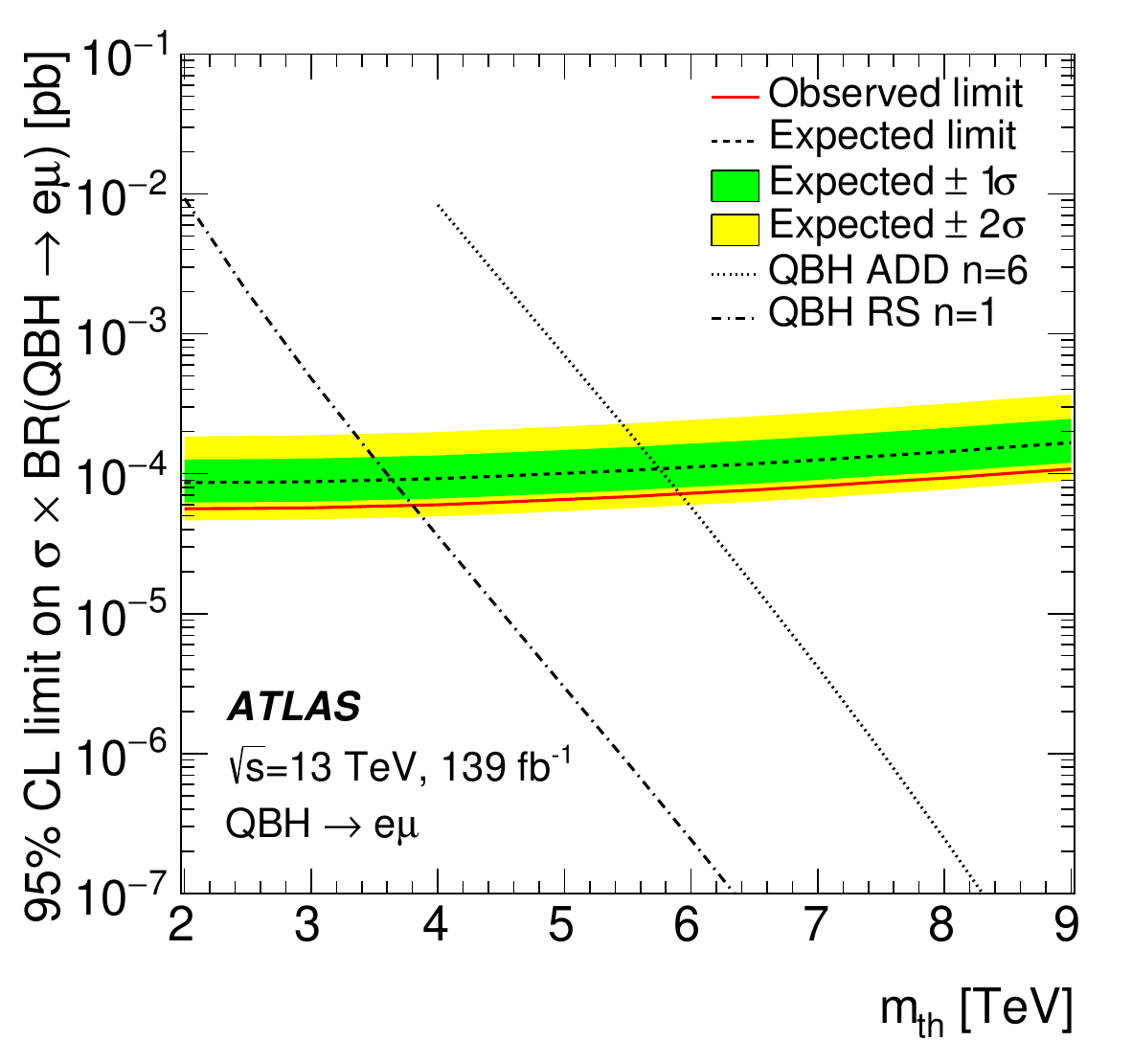}}
\caption{The observed and expected 95\% CL upper limits on the
(a) Z$^\prime$ boson, (b) RPV $\tau$-sneutrino ($\tilde{\nu}_{\tau}$) and (c) QBH
ADD and RS production cross-section times branching ratio for decays into an
$e\mu$ final state. The signal
theoretical cross-section times branching ratio lines for the \Zprime model,
the QBH ADD model assuming six extra dimensions, and the RS model with one
extra dimension are obtained from the simulation of each
process, while the \Zprime is corrected to NNLO and the RPV $\tilde{\nu}_{\tau}$ is corrected to NLO. The theoretical uncertainties are not considered in the mass limit calculation. The acceptance times
efficiency of the ADD and RS QBH models agree to within 1\% and the same
curve is used for limit extraction. The expected limits are shown with the
$\pm 1$ and $\pm 2$ standard deviation uncertainty band.}
\label{fig:limit_emu}
\end{figure}

\begin{figure}[!htbp]
\centering
\subfloat[]{
\includegraphics[width=0.49\linewidth]{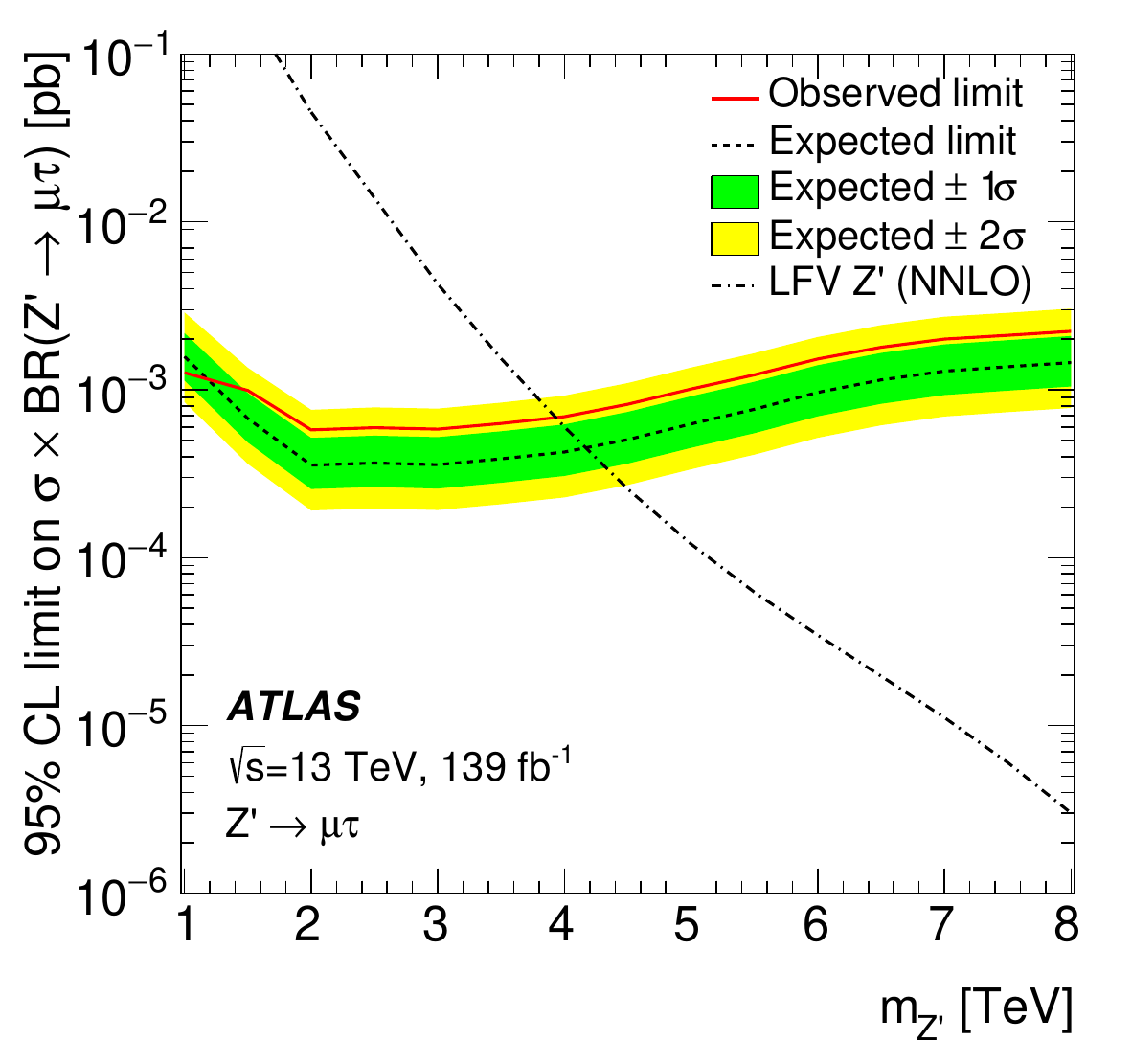}}
\subfloat[]{
\includegraphics[width=0.49\linewidth]{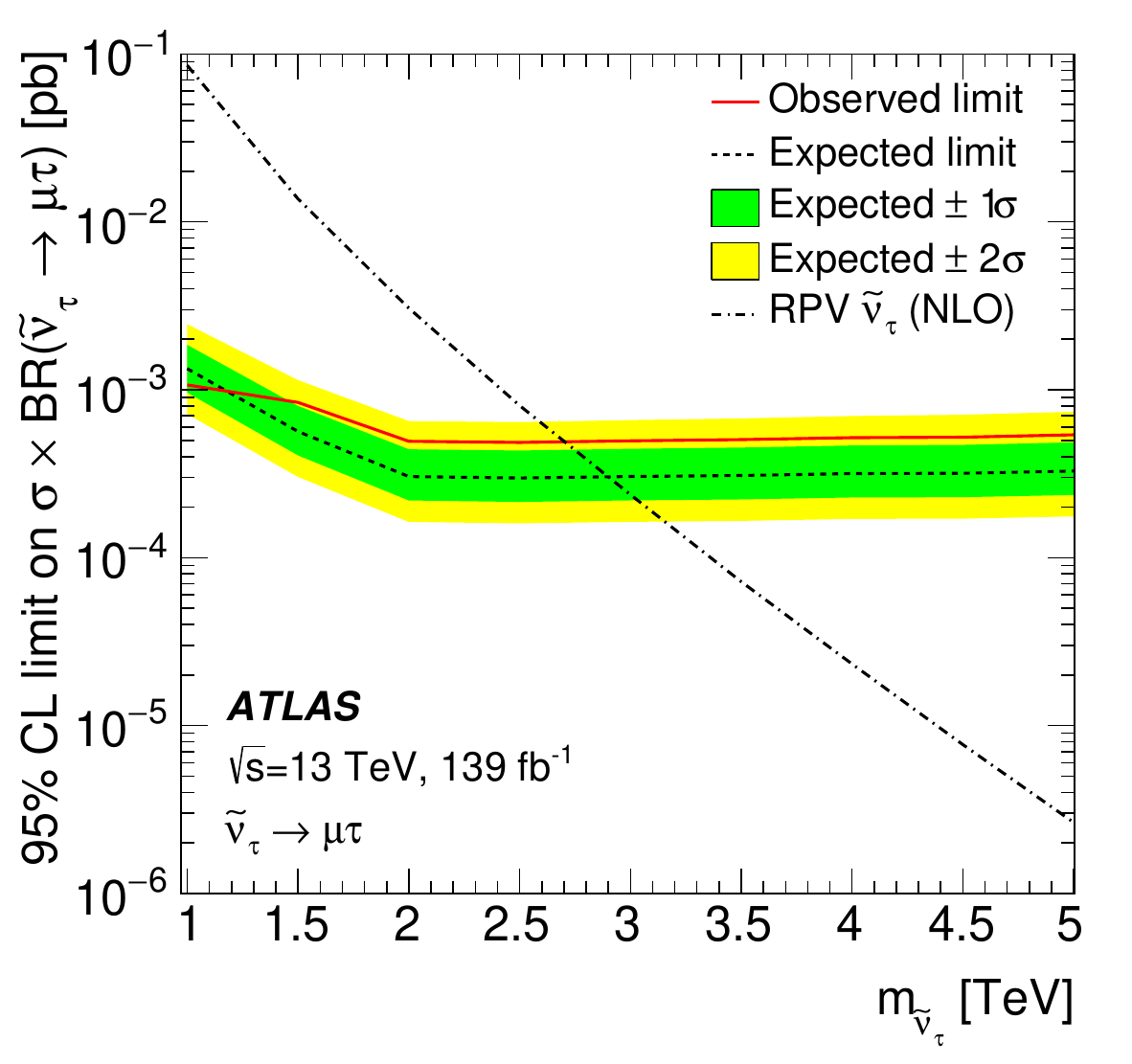}} \\
\subfloat[]{
\includegraphics[width=0.49\linewidth]{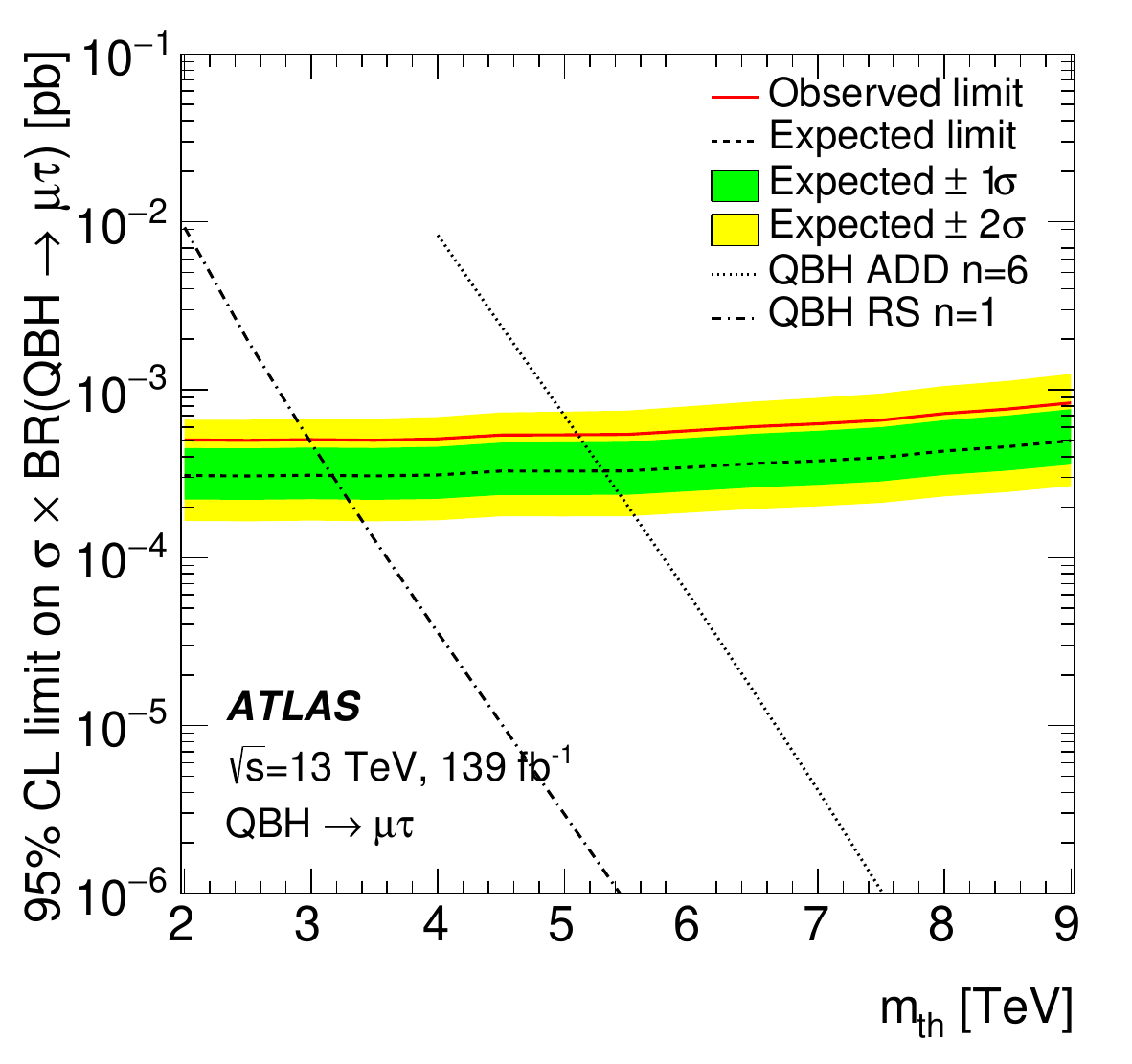}}
\caption{The observed and expected 95\% CL upper limits on the
(a) \Zprime boson, (b) RPV $\tau$-sneutrino ($\tilde{\nu}_{\tau}$) and (c) QBH
ADD and RS production cross-section times branching ratio for decays into a
$\mu\tau$ final state. The signal theoretical cross-section times branching
ratio lines for the \Zprime model, the QBH ADD model assuming six extra
dimensions, and the RS model with one extra dimension are obtained from the
simulation of each process, while the \Zprime is corrected to NNLO and the RPV $\tilde{\nu}_{\tau}$ is corrected to NLO. The theoretical uncertainties are not considered in the mass limit calculation. The acceptance times efficiency of the ADD and RS QBH
models agree to within 1\% and the same curve is used for limit
extraction. The expected limits are shown with the $\pm 1$ and $\pm 2$
standard deviation uncertainty bands.}
\label{fig:limit_mutau}
\end{figure}

Figures~\ref{fig:limit_emu}--\ref{fig:limit_mutau} show the observed and expected 95\% CL upper limits on the production cross-section times branching ratio of the \Zprime, RPV SUSY $\tilde{\nu}_{\tau}$ and QBH models for each of the final states considered. The limits for signal masses above 3.0 \TeV are dominated by the last bin in the dilepton invariant mass spectrum, thus the limit curve tends to be flat in the high mass region. 
 
The extracted limits are not as strong for signal masses above about 2.5 \TeV due to a decrease in acceptance at very high $p_{\mathrm{T}}$, and specifically to the LFV \Zprime model, low-mass signal production due to PDF suppression at high masses. 
The observed limits are slightly more stringent than the expected limits above 2 \TeV because of the small deficit observed in data in $e\mu$ channel.
In the $\tau$-lepton channels, the observed limits are less stringent than the expected limits above 2 \TeV because of the mild excess above
the background estimates between 2.0 \TeV and 2.3 \TeV. The acceptance times
efficiency of the ADD and RS QBH models agree with each other within 1\%, and the same prediction is used for the limit extraction. The results are summarised in Table~\ref{tab:limits}.

\begin{table}[htbp]
\begin{center}
\caption[]{Observed data event yields and the expected background event yields with their total uncertainties in the CRs in the $e\mu$ channel after applying all selection criteria and the background-only fit. The ``Fake'' background refers to ``\W+jets'' and ``Multijet'' events with fake leptons. The ``Top Quarks'' contains single top and $t\bar{t}$ events. }
\label{Tab:CR_emu}\vspace{12pt}
\begin{tabular}{l|c|c|c}
\toprule
Process    & {\ttbar CR}      & {$WW$ CR}     & {Low $\Delta\phi_{\ell\ell'}$ \ttbar CR}    \\
\midrule
Top Quarks & 660$\pm$27    & 55.4$\pm$5.9    & 279$\pm$18   \\
Diboson    & 11.3$\pm$2.4    & 54.3$\pm$8.4    & 7.5$\pm$2.3    \\
Fake       & 29.3$\pm$5.9    & 18.3$\pm$3.8   & 16.6$\pm$4.9   \\
Drell--Yan  & 0.25$\pm$0.03     & 0.26$\pm$0.03    & 0    \\
\midrule
Total background & 701$\pm$26    & 128.2$\pm$8.3  & 304$\pm$17  \\
Data       & 700     & 133     & 301    \\
\bottomrule
\end{tabular}
\end{center}
\end{table}
 
\begin{table}[htbp]
\begin{center}
\caption[]{Observed data event yields and the expected background event yields with their total uncertainties in the CRs in the $e\tau$ and $\mu\tau$ channels after applying all selection criteria and the background-only fit. The ``Fake'' background is estimated directly by \W+jets MC, and the multijet contribution is not considered in the \ttbar CRs. The ``Top Quarks'' contains single top and $t\bar{t}$ events.}
\label{Tab:CR_tau}\vspace{12pt}
\begin{tabular}{l|c|c}
\toprule
Process    & {$e\tau$ channel \ttbar CR}      & {$\mu\tau$ channel \ttbar CR}    \\
\midrule
Top Quarks & 406$\pm$21    & 305$\pm$17     \\
Diboson    & 9.6$\pm$2.8    & 6.1$\pm$1.6       \\
Fake       & 6.8$\pm$1.2    & 8.6$\pm$2.6      \\
Drell--Yan  & 2.1$\pm$0.1     & 3.6$\pm$0.7      \\
\midrule
Total background & 424$\pm$20    & 324$\pm$17  \\
Data       & 422     & 324         \\
\bottomrule
\end{tabular}
\end{center}
\end{table}
 
\begin{table}[htbp]
\begin{center}
\caption[]{Observed data event yields and the expected background event yields with their total uncertainties in the SR ($m_{\ell\ell'} > 600$ \si{GeV}) in the $e\mu$, $e\tau$, and $\mu\tau$ channels after applying all selection criteria and the background-only fit. The ``Fake'' background refers to ``\W+jets'' and ``Multijet'' events with fake leptons. The ``Top Quarks'' contains single top and $t\bar{t}$ events.}
\label{Tab:SR_emu_etau_mutau}\vspace{12pt}
\begin{tabular}{l|c|c|c}
\toprule
Process    & {$e\mu$ channel}      & {$e\tau$ channel}     & {$\mu\tau$ channel}    \\
\midrule
Top Quarks & 151$\pm$15   & 114$\pm$10    & 79.4$\pm$6.4   \\
Diboson    & 246$\pm$28   & 125$\pm$27    & 94$\pm$20    \\
Fake       & 66$\pm$11    & 172$\pm$34    & 67$\pm$25   \\
Drell--Yan  & 8.6$\pm$0.5      & 76.1$\pm$8.9     & 78.0$\pm$7.9    \\
\midrule
Total background   & 471$\pm$21   &  488$\pm$21  & 319$\pm$16  \\
Data       & 470     & 499     & 319    \\
\bottomrule
\end{tabular}
\end{center}
\end{table}
 
\begin{table}[htbp]
\begin{center}
\caption[]{The fitted correction factors of ``Top Quarks'' and ``Diboson'' backgrounds which are applied in the SR for the background-only fit in the $e\mu$, $e\tau$ and $\mu\tau$ channel. For the $e\tau$ and $\mu\tau$ channels, the diboson correction factors are calculated from the CR-only fit results and put into the final fit as fixed normalisation factors.} 
\label{Tab:NormF}\vspace{12pt}
\begin{tabular}{c|c|c|c}\hline \hline
Correction factor	& $e\mu$     & $e\tau$    & $\mu\tau$    \\
\hline
Top Quarks      & 0.89$\pm0.06$     &	0.94$\pm0.11$ & 0.97$\pm0.11$       \\
Diboson   & 0.94$\pm0.11$     &	1.05$\pm0.12$ & 1.00$\pm0.12$        \\
\hline\hline
 
\end{tabular}
\end{center}
\end{table}
 
\begin{figure}[!htbp]
\centering
\subfloat[]{
\includegraphics[width=0.49\linewidth]{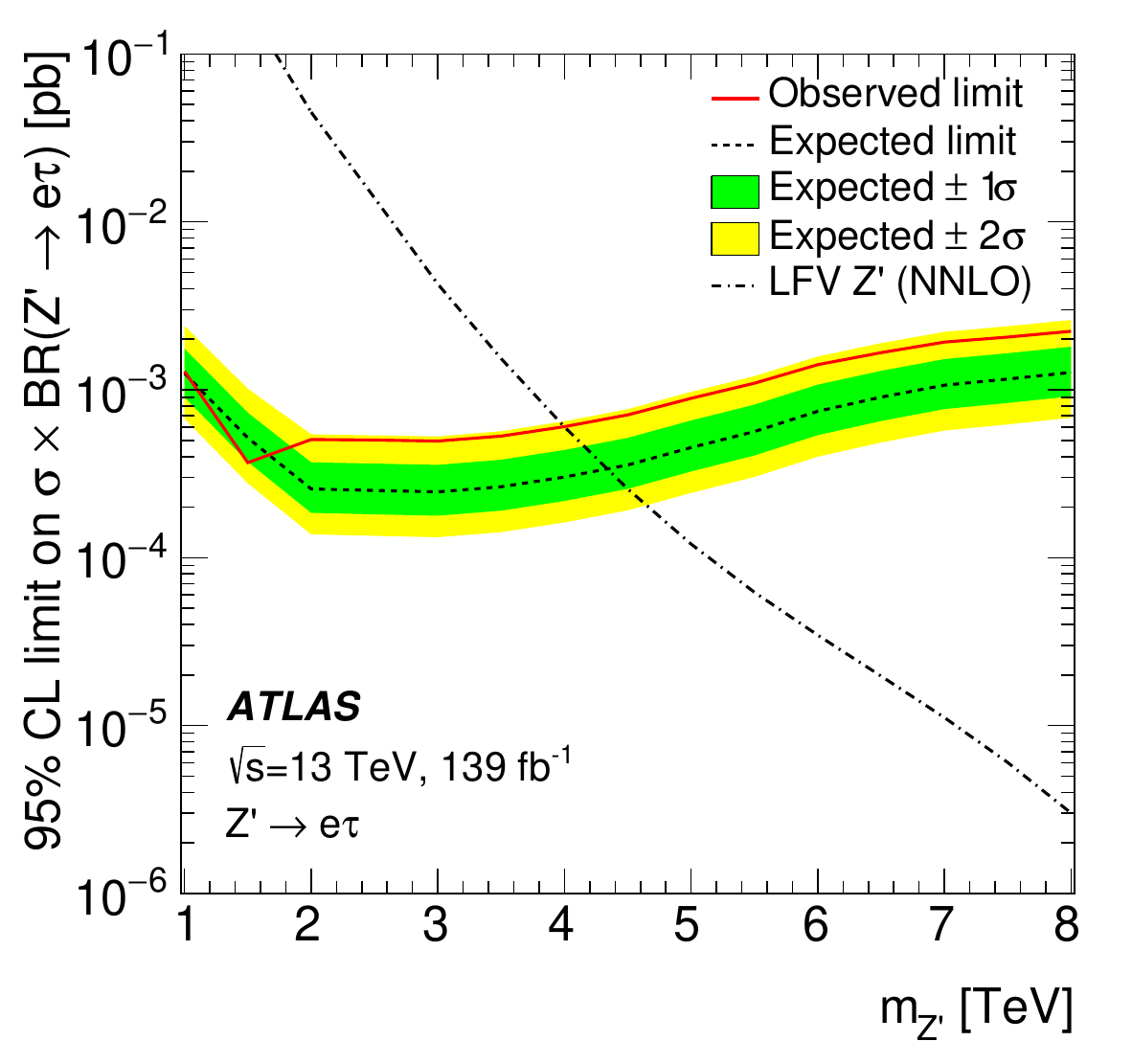}}
\subfloat[]{
\includegraphics[width=0.49\linewidth]{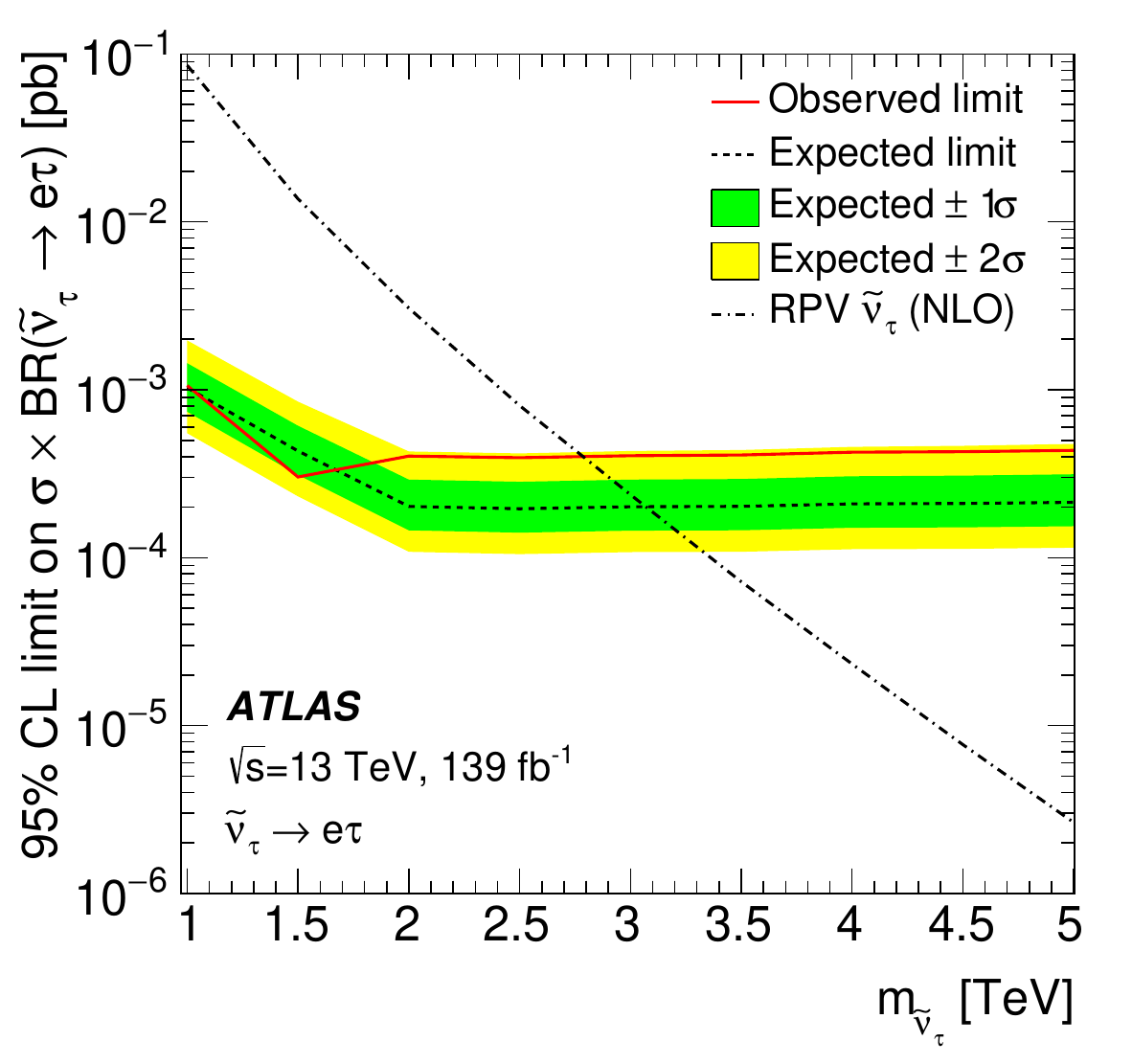}} \\
\subfloat[]{
\includegraphics[width=0.49\linewidth]{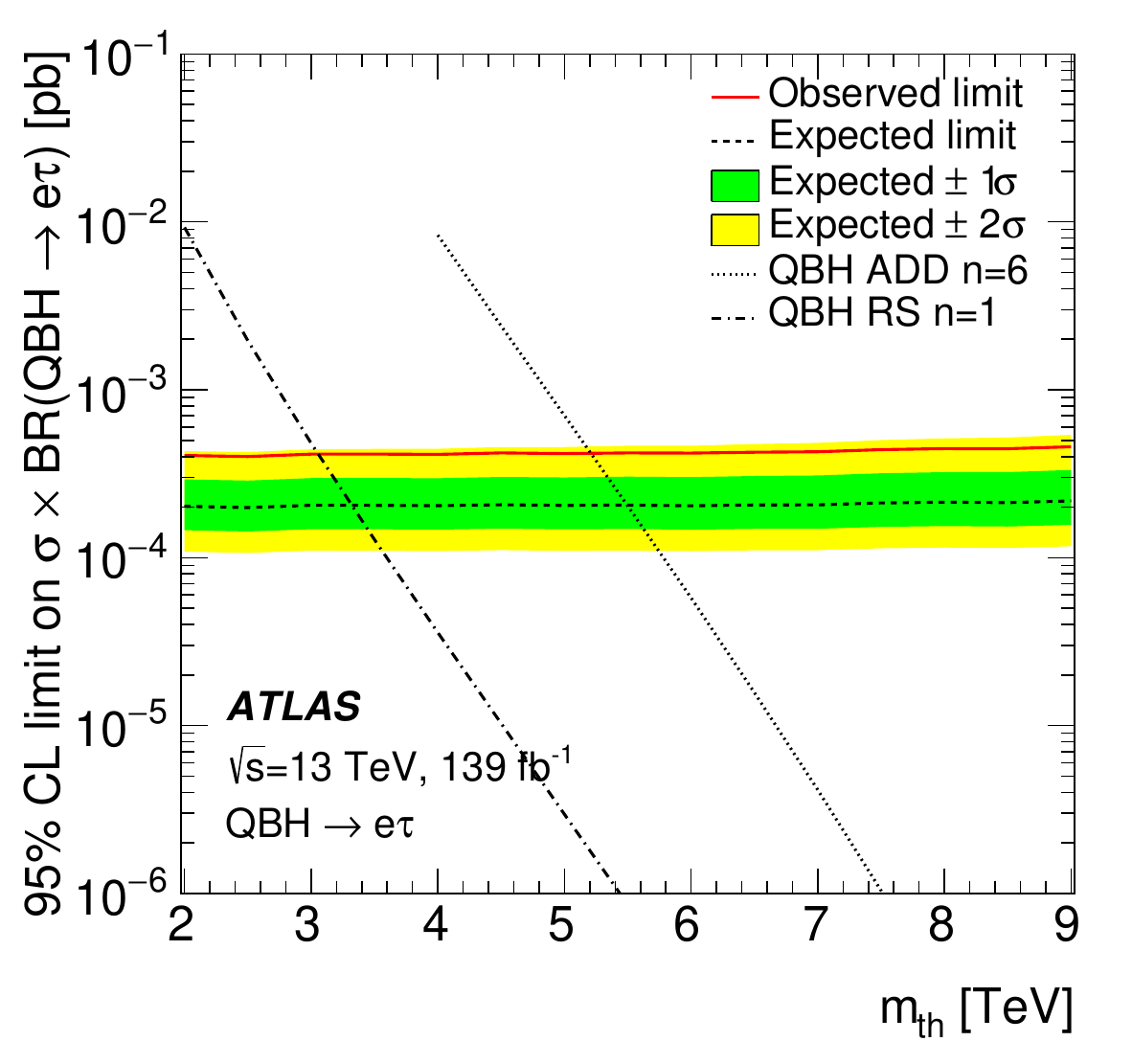}}
\caption{The observed and expected 95\% CL upper limits on the
(a) \Zprime boson, (b) RPV $\tau$-sneutrino ($\tilde{\nu}_{\tau}$) and (c) QBH
ADD and RS production cross-section times branching ratio for decays into an
$e\tau$ final state. The signal theoretical cross-section times branching
ratio lines for the \Zprime model, the QBH ADD model assuming six extra
dimensions, and the RS model with one extra dimension are obtained from the
simulation of each process, while the \Zprime is corrected to NNLO and the RPV $\tilde{\nu}_{\tau}$ is corrected to NLO. The theoretical uncertainties are not considered in the mass limit calculation. The acceptance times efficiency of the ADD and RS QBH
models agree to within 1\% and the same curve is used for limit
extraction. The expected limits are shown with the $\pm 1$ and $\pm 2$
standard deviation uncertainty bands.}
\end{figure}

\begin{table}[!htbp]
\centering
\caption{Expected and observed 95\% CL lower limits on the
mass of a \Zprime boson with lepton-flavour-violating couplings, a
supersymmetric $\tau$-sneutrino ($\tilde{\nu}_{\tau}$) with
$R$-parity-violating couplings, and the threshold mass for quantum black-hole
production for the ADD $n=6$ and RS $n=1$ models.}
\begin{tabular}{l|ccc}
\toprule
Model
& \multicolumn{3}{c}{Observed (expected) 95\% CL lower limit [\si{TeV}]}\\
\cline{2-4}
& $e\mu$ channel & $e\tau$ channel & $\mu\tau$ channel\\
\midrule
LFV \Zprime
& 5.0 (4.8) & 4.0 (4.3) & 3.9 (4.2)\\
RPV SUSY $\tilde{\nu}_{\tau}$
& 3.9 (3.7) & 2.8 (3.0) & 2.7 (2.9)\\
QBH ADD $n=6$
& 5.9 (5.7) & 5.2 (5.5) & 5.1 (5.2)\\
QBH RS $n=1$
& 3.8 (3.6) & 3.0 (3.3) & 3.0 (3.1)\\
\bottomrule
\end{tabular}
\label{tab:limits}
\end{table}


\FloatBarrier
 
\section{Conclusions}

A search for a heavy particle decaying into $e\mu$, $e\tau$ or $\mu\tau$ final states is conducted using 139 fb$^{-1}$ of proton--proton collision data at $\sqrt{s}=13$ \TeV recorded by the ATLAS detector at the Large Hadron Collider. The data are found to be consistent with the Standard Model expectation in 
the SR ($m_{\ell\ell'} > 600$ \GeV). With no evidence for new physics, profile likelihood fits are used to
set 95\% CL lower limits on the mass of a Z$^\prime$ vector boson with lepton-flavour-violating couplings
at 5.0, 4.0 and 3.9 \TeV for the $e\mu$, $e\tau$ or $\mu\tau$ final states, respectively; on the mass of a supersymmetric $\tau$-sneutrino with
R-parity-violating couplings at 3.9, 2.8 and 2.7 \TeV; and on the threshold mass for quantum black-hole
production in the context of the ADD n=6 (RS n=1) model at 5.9 (3.8),
5.2 (3.0) and 5.1 (3.0) \TeV respectively.
 
For the mass limit of the \Zprime signal model, as an illustration, the observed results based on the full data samples recorded by
the ATLAS experiment at $\sqrt{s}=13$ \TeV are more stringent by $0.6$, $0.3$ and $0.4$ TeV than the
corresponding limits based on the \SI{36.1}{\per\femto\barn} data samples at $\sqrt{s}=13$ \TeV~\cite{EXOT-2016-09}.
The expected sensitivity is improved by $0.5$, $0.6$ and $0.7$ \TeV for the $e\mu$, $e\tau$ and $\mu\tau$
channels respectively. The increases on the observed limits in $e\tau$ and $\mu\tau$ channels are not as strong as the expected ones due to the slight excess observed in data. The major improvement comes from the four times larger data sample.
The increase of sensitivity is also due to the application of b-jet veto, more accurate data-driven background estimates, as well as better particle reconstruction and identification.


\clearpage
 
\section*{Acknowledgements}


We thank CERN for the very successful operation of the LHC, as well as the
support staff from our institutions without whom ATLAS could not be
operated efficiently.
 
We acknowledge the support of
ANPCyT, Argentina;
YerPhI, Armenia;
ARC, Australia;
BMWFW and FWF, Austria;
ANAS, Azerbaijan;
CNPq and FAPESP, Brazil;
NSERC, NRC and CFI, Canada;
CERN;
ANID, Chile;
CAS, MOST and NSFC, China;
Minciencias, Colombia;
MEYS CR, Czech Republic;
DNRF and DNSRC, Denmark;
IN2P3-CNRS and CEA-DRF/IRFU, France;
SRNSFG, Georgia;
BMBF, HGF and MPG, Germany;
GSRI, Greece;
RGC and Hong Kong SAR, China;
ISF and Benoziyo Center, Israel;
INFN, Italy;
MEXT and JSPS, Japan;
CNRST, Morocco;
NWO, Netherlands;
RCN, Norway;
MEiN, Poland;
FCT, Portugal;
MNE/IFA, Romania;
MESTD, Serbia;
MSSR, Slovakia;
ARRS and MIZ\v{S}, Slovenia;
DSI/NRF, South Africa;
MICINN, Spain;
SRC and Wallenberg Foundation, Sweden;
SERI, SNSF and Cantons of Bern and Geneva, Switzerland;
MOST, Taiwan;
TENMAK, T\"urkiye;
STFC, United Kingdom;
DOE and NSF, United States of America.
In addition, individual groups and members have received support from
BCKDF, CANARIE, Compute Canada and CRC, Canada;
PRIMUS 21/SCI/017 and UNCE SCI/013, Czech Republic;
COST, ERC, ERDF, Horizon 2020 and Marie Sk{\l}odowska-Curie Actions, European Union;
Investissements d'Avenir Labex, Investissements d'Avenir Idex and ANR, France;
DFG and AvH Foundation, Germany;
Herakleitos, Thales and Aristeia programmes co-financed by EU-ESF and the Greek NSRF, Greece;
BSF-NSF and MINERVA, Israel;
Norwegian Financial Mechanism 2014-2021, Norway;
NCN and NAWA, Poland;
La Caixa Banking Foundation, CERCA Programme Generalitat de Catalunya and PROMETEO and GenT Programmes Generalitat Valenciana, Spain;
G\"{o}ran Gustafssons Stiftelse, Sweden;
The Royal Society and Leverhulme Trust, United Kingdom.
 
The crucial computing support from all WLCG partners is acknowledged gratefully, in particular from CERN, the ATLAS Tier-1 facilities at TRIUMF (Canada), NDGF (Denmark, Norway, Sweden), CC-IN2P3 (France), KIT/GridKA (Germany), INFN-CNAF (Italy), NL-T1 (Netherlands), PIC (Spain), ASGC (Taiwan), RAL (UK) and BNL (USA), the Tier-2 facilities worldwide and large non-WLCG resource providers. Major contributors of computing resources are listed in Ref.~\cite{ATL-SOFT-PUB-2023-001}.


\clearpage
\printbibliography
\clearpage
 
\clearpage
\input{atlas_authlist}

\clearpage

\end{document}

%% file: atlas_authlist.tex
 
\begin{flushleft}
\hypersetup{urlcolor=black}
{\Large The ATLAS Collaboration}

\bigskip

\AtlasOrcid[0000-0002-6665-4934]{G.~Aad}$^\textrm{\scriptsize 102}$,
\AtlasOrcid[0000-0002-5888-2734]{B.~Abbott}$^\textrm{\scriptsize 120}$,
\AtlasOrcid[0000-0002-1002-1652]{K.~Abeling}$^\textrm{\scriptsize 55}$,
\AtlasOrcid[0000-0001-5763-2760]{N.J.~Abicht}$^\textrm{\scriptsize 49}$,
\AtlasOrcid[0000-0002-8496-9294]{S.H.~Abidi}$^\textrm{\scriptsize 29}$,
\AtlasOrcid[0000-0002-9987-2292]{A.~Aboulhorma}$^\textrm{\scriptsize 35e}$,
\AtlasOrcid[0000-0001-5329-6640]{H.~Abramowicz}$^\textrm{\scriptsize 151}$,
\AtlasOrcid[0000-0002-1599-2896]{H.~Abreu}$^\textrm{\scriptsize 150}$,
\AtlasOrcid[0000-0003-0403-3697]{Y.~Abulaiti}$^\textrm{\scriptsize 117}$,
\AtlasOrcid[0000-0003-0762-7204]{A.C.~Abusleme~Hoffman}$^\textrm{\scriptsize 137a}$
% ...
\bigskip
\\

% ...
$^{29}$Physics Department, Brookhaven National Laboratory, Upton NY; United States of America.\\
% ...
$^{35}$$^{(a)}$Facult\'e des Sciences Ain Chock, R\'eseau Universitaire de Physique des Hautes Energies - Universit\'e Hassan II, Casablanca;$^{(b)}$Facult\'{e} des Sciences, Universit\'{e} Ibn-Tofail, K\'{e}nitra;$^{(c)}$Facult\'e des Sciences Semlalia, Universit\'e Cadi Ayyad, LPHEA-Marrakech;$^{(d)}$LPMR, Facult\'e des Sciences, Universit\'e Mohamed Premier, Oujda;$^{(e)}$Facult\'e des sciences, Universit\'e Mohammed V, Rabat;$^{(f)}$Institute of Applied Physics, Mohammed VI Polytechnic University, Ben Guerir; Morocco.\\
% ...
$^{49}$Fakult\"{a}t Physik , Technische Universit{\"a}t Dortmund, Dortmund; Germany.\\
% ...
$^{55}$II. Physikalisches Institut, Georg-August-Universit\"{a}t G\"ottingen, G\"ottingen; Germany.\\
% ...
$^{102}$CPPM, Aix-Marseille Universit\'e, CNRS/IN2P3, Marseille; France.\\
% ...
$^{117}$Department of Physics, New York University, New York NY; United States of America.\\
% ...
$^{120}$Homer L. Dodge Department of Physics and Astronomy, University of Oklahoma, Norman OK; United States of America.\\
% ...
$^{137}$$^{(a)}$Departamento de F\'isica, Pontificia Universidad Cat\'olica de Chile, Santiago;$^{(b)}$Millennium Institute for Subatomic physics at high energy frontier (SAPHIR), Santiago;$^{(c)}$Instituto de Investigaci\'on Multidisciplinario en Ciencia y Tecnolog\'ia, y Departamento de F\'isica, Universidad de La Serena;$^{(d)}$Universidad Andres Bello, Department of Physics, Santiago;$^{(e)}$Instituto de Alta Investigaci\'on, Universidad de Tarapac\'a, Arica;$^{(f)}$Departamento de F\'isica, Universidad T\'ecnica Federico Santa Mar\'ia, Valpara\'iso; Chile.\\
% ...
$^{150}$Department of Physics, Technion, Israel Institute of Technology, Haifa; Israel.\\
$^{151}$Raymond and Beverly Sackler School of Physics and Astronomy, Tel Aviv University, Tel Aviv; Israel.\\
% ...
$^{a}$ Also Affiliated with an institute covered by a cooperation agreement with CERN.\\
$^{b}$ Also at An-Najah National University, Nablus; Palestine.\\
$^{c}$ Also at APC, Universit\'e Paris Cit\'e, CNRS/IN2P3, Paris; France.\\
$^{d}$ Also at Borough of Manhattan Community College, City University of New York, New York NY; United States of America.\\
$^{e}$ Also at Center for High Energy Physics, Peking University; China.\\
$^{f}$ Also at Center for Interdisciplinary Research and Innovation (CIRI-AUTH), Thessaloniki; Greece.\\
% ...

\end{flushleft}
